\documentclass[twocolumn,         % Format : preprint, twocolumn
               showpacs,longbibliography,            % Pacs : showpacs, noshowpacs
               showkeys,preprintnumbers,     % Preprint: preprintnumbers,
               aps,                 % Society: ...
               prd,          	    % Journal Style : pra, prb, prc, prd, pre,
               letterpaper,             % Size : a4paper, ...
               superscriptaddress,      % Affiliation (Title) : groupedaddress,
               nofootinbib,         % Footnote: footinbib, nofootinbib
               tightenlines,        % Remove additional spaces in a line
               floats,floatfix      % Floating pictures and tables
               ]{revtex4-1}
\usepackage{physics}
\usepackage{epsf}
\usepackage{graphicx,color}
\usepackage{subfigure}
\usepackage{latexsym}
\usepackage{amsmath,amssymb}        
\usepackage[colorlinks=true,linkcolor=blue,citecolor=blue]{hyperref}
\usepackage{mathrsfs}
\usepackage{comment}
\usepackage{soul}
\usepackage{cancel}
\definecolor{purple}{rgb}{0.58,0.0,0.83}
\definecolor{orange}{rgb}{1,0.5,0}
\DeclareSymbolFontAlphabet{\mathrsfs}{rsfs}
\DeclareMathAlphabet{\mathcal}{OMS}{cmsy}{m}{n}

\begin{document}

% -----> TITLE 

\title{Time-dependent adaptive mesh refinement solver for the Gross-Pitaevskii-Poisson equations}

% ----->   AUTHORS   <-----

\author{Iv\'an  \'Alvarez-Rios}
\email{ivan.alvarez@umich.mx}
\affiliation{Instituto de F\'{\i}sica y Matem\'{a}ticas, Universidad
              Michoacana de San Nicol\'as de Hidalgo. Edificio C-3, Cd.
              Universitaria, 58040 Morelia, Michoac\'{a}n,
              M\'{e}xico.}

% --->   DATE

\date{\today}

% -----> ABSTRACT

\begin{abstract}
This work presents a new numerical code for solving the time--dependent Gross--Pitaevskii--Poisson (GPP) system using adaptive mesh refinement (AMR). The code is designed to study the nonlinear dynamics of self--gravitating bosonic matter in three spatial dimensions under periodic boundary conditions. It combines high--order spatial discretization, explicit time integration, and dynamic refinement driven by the magnitude of the gravitational potential. The implementation is validated through a set of test problems in the nonlinear regime. These benchmarks demonstrate that the solver accurately preserves global conservation laws, resolves strong wave interference and phase singularities, and maintains consistency across refinement levels in highly dynamical scenarios.
\end{abstract}

% ----->   PACS

% ----->   MAKETITLE   <-----

\maketitle
%-------------------> SEC: introduction
\section{Introduction}
\label{sec:intro}

One of the most extensively studied alternatives to the standard Cold Dark Matter (CDM) paradigm~\cite{Peebles:1982} is Bose--Einstein Condensate Dark Matter (BECDM), also commonly referred to as Fuzzy Dark Matter (FDM), ultralight bosonic dark matter, or scalar--field dark matter, depending on the boson mass and self--interaction properties \cite{Matos-Urena:2000,GuzmanUrena2004,Schive:2014hza,Hui:2016}. In this framework, dark matter is composed of extremely light bosons whose large de Broglie wavelength gives rise to macroscopic quantum phenomena on galactic and subgalactic scales.

The wave nature of BECDM gives rise to distinctive quantum phenomena at astrophysical scales. At linear order, the quantum description introduces a cutoff in the matter power spectrum, suppressing structure formation below a characteristic scale determined by the boson mass~\cite{Hu:2000}. At nonlinear scales, quantum interference effects generate an effective pressure—commonly referred to as quantum pressure—that counteracts gravitational collapse in the inner regions of halos. This mechanism naturally leads to the emergence of stable, self--gravitating solitonic cores at galactic centers~\cite{Schive:2014dra}, embedded within extended halos arising from the coarse--grained dynamics of the interference pattern.

Over the past two decades, the BECDM framework has been investigated from multiple complementary perspectives, ranging from equilibrium configurations to large--scale numerical simulations of structure formation \cite{GuzmanUrena2004,AlvarezGuzmanMadelung,AlvarezGuzman2022,CarlosIvanFranciscoUniverse,Schive:2014dra,Mocz:2017wlg,mocz19b}. Early studies focused on stationary solutions of the Schr\"odinger--Poisson (SP) system, identifying self--gravitating solitonic configurations that can be interpreted as galactic cores or nuclei~\cite{Matos-Urena:2000,GuzmanUrena2004,AlvarezGuzmanMadelung,CarlosIvanFranciscoUniverse}. These solutions have subsequently been extended to include additional physical ingredients, such as central black holes~\cite{Moczfdmbh,palomareschavez2025} and baryonic components modeled as ideal or polytropic fluids~\cite{AlvarezGuzman2022,AlvarezGuzmanNiemeyer_2025}, enabling more realistic descriptions of galactic environments.

Beyond isolated equilibria, the dynamical interaction of BECDM structures has been studied extensively. Binary mergers of solitonic cores exhibit a rich phenomenology dominated by wave interference, gravitational cooling, and relaxation toward final bound states~\cite{Schive:2014hza,GuzmanAlvarezGonzalez2021,AlvarezGuzman2022,periodicas}. These analyses have been extended to multi--core mergers and cluster--like configurations, providing insight into the formation of core--halo structures and the scaling relations linking core and halo properties~\cite{Schive:2014dra,Mocz:2017wlg,periodicas}.

At larger scales, numerical simulations have shown that BECDM can reproduce cosmic structure formation while suppressing small--scale power, both in fully cosmological settings and in controlled local environments~\cite{Hu:2000,Hui:2016}. These studies have clarified the transition from the wave--dominated regime at small scales to an effectively classical behavior at larger scales, where coarse--grained quantities resemble those of collisionless CDM.

In addition to these scenarios, BECDM admits a variety of nontrivial solutions associated with its complex scalar nature. These include vortex lines, vortex lattices, and other topological defects arising from phase singularities, which have been proposed as signatures of rotation and angular momentum transport in galactic halos~\cite{Nikolaieva_2021,Nikolaieva_2023,Glennon2023,Rindler-Shapiro:2012,Alvarez_Rios_2025}.

From a computational perspective, these advances have been enabled by a wide range of numerical approaches. Several groups have developed dedicated simulation codes based on finite--difference, spectral, or pseudo--spectral methods, often incorporating adaptive mesh refinement (AMR) to resolve the large dynamical range inherent to BECDM dynamics~\cite{Schive:2014hza,Schive:2014dra,Mocz:2017wlg}. Among the most widely used frameworks are {\tt Nyx}, which employs a grid--based AMR approach for cosmological structure formation including scalar field dark matter~\cite{Schive:2014hza}, and {\tt GAMER}, a GPU--accelerated AMR code extended to solve the SP system in cosmological settings~\cite{Schive:2010dd,Schive:2014dra}. Moving--mesh codes such as {\tt AREPO}~\cite{Springel:2010ux} have also been adapted to incorporate wave--like dark matter dynamics, while AMR frameworks such as {\tt Enzo}~\cite{Bryan:2014vza} provide flexible infrastructures for cosmological simulations.

Complementary to these large--scale solvers, several specialized implementations have been developed to directly evolve the SP or Gross--Pitaevskii--Poisson (GPP) systems using pseudo--spectral or high--order finite--difference methods. Examples include codes such as {\tt PyUltraLight}~\cite{Edwards:2018lsn}, designed to accurately capture solitonic cores, interference patterns, and vortex dynamics at galactic and subgalactic scales.

The existence of multiple independently developed and validated codes is essential to ensure the reliability of numerical results in BECDM simulations. Given the complexity of the underlying equations and the diversity of numerical approaches, cross--validation between independent implementations plays a crucial role in identifying systematic effects and strengthening confidence in physical predictions. Despite this progress, constructing numerical frameworks that are simultaneously robust, flexible, and well validated remains a significant challenge, particularly when addressing long--term stability, conservation properties, and parent--child consistency in AMR simulations.

In this work, we present a new implementation of the \texttt{CAFE--FDM} code designed to solve the GPP system using adaptive mesh refinement. This development builds upon earlier versions of the code that have been successfully applied in different contexts. The first version employed fixed mesh refinement and isolated boundary conditions, using explicit time integrators and finite--difference schemes~\cite{AlvarezGuzman2022}. A second version introduced periodic boundary conditions and pseudo--spectral methods~\cite{periodicas}, enabling the study of wave--dominated regimes. These implementations have been used to investigate a variety of BECDM scenarios, including the stability of core--tail structures~\cite{IvanFranciscoCoreTailSols}, mergers of galactic cores~\cite{GuzmanAlvarezGonzalez2021}, fermion--boson stars~\cite{Alvarez_Rios_2023,AlvarezGuzmanNiemeyer_2025}, boson stars around black holes~\cite{palomareschavez2025}, and dark--matter accretion in low--surface--brightness galaxies~\cite{Alvarez_Rios_2024}.

Despite these advances, previous versions relied on fixed mesh refinement or uniform grids, which become inefficient when addressing problems involving a wide range of dynamically relevant scales. In BECDM systems, highly localized structures---such as solitonic cores and vortex filaments---coexist with extended interference patterns and large--scale gravitational dynamics. Accurately resolving these features requires a strategy that concentrates computational resources where they are most needed.

Adaptive mesh refinement provides a natural framework to address this multiscale behavior by dynamically allocating resolution in regions of interest. This allows small--scale features, such as phase singularities and steep gradients, to be resolved while preserving global structures and reducing computational cost.

Beyond the scalar--field formulation, the \texttt{CAFE--FDM} framework is designed with sufficient generality to accommodate multi--component wavefunctions. This capability was exploited in Ref.~\cite{Alvarez_Guzman_2025}, where stationary and dynamical solutions of the SP system for $n$--dimensional states were constructed. While the present work focuses on the scalar ($n=1$) case as a baseline for validating the AMR implementation, the underlying architecture can be naturally extended to multi--field configurations.

The present work describes this new implementation, providing an independent AMR framework that combines adaptive refinement with systematic validation in both unigrid and refined settings. The code is validated through a comprehensive suite of tests, including stationary configurations, boosted solutions, strongly nonlinear mergers, wave--dominated regimes, and the formation of solitonic attractors through gravitational Bose--Einstein condensation. As such, it provides a reliable and extensible tool for future investigations of BECDM phenomenology and related self--gravitating quantum systems.

This paper is organized as follows. In Sec.~\ref{sec:model} we introduce the theoretical framework and governing equations. Section~\ref{sec:AMR} describes the numerical methods and adaptive mesh refinement strategy. In Sec.~\ref{sec:validation} we present the validation tests and results. Finally, Sec.~\ref{sec:conclusions} summarizes our conclusions and outlines future directions.
%-------------------> SEC: model and equations
\section{Model and equations}
\label{sec:model}

\subsection{Gross--Pitaevskii--Poisson system}
\label{sec:gpp}

The dynamics of BECDM is governed by the GPP system, written in physical units and Cartesian coordinates $(\tilde t,\vec{\tilde x})$ \citep{GuzmanUrena2004,GuzmanUrena2006,AlvarezGuzman2022,Mocz-Robles:2017}:

\begin{align}
    {\rm i}\hbar\,\partial_{\tilde t}\tilde\Psi
    &=\Big[-\tfrac{\hbar^2}{2m_B}\tilde\nabla^2
      + m_B\,\tilde V
      + \tilde g\,|\tilde\Psi|^2\Big]\tilde\Psi,
    \label{eq:gross-pitaevskii}\\[3pt]
    \tilde\nabla^2 \tilde V
    &=4\pi G\,(\tilde\rho-\bar{\tilde\rho}),
    \label{eq:poisson}
\end{align}

\noindent where $\tilde\Psi$ is the macroscopic wavefunction, namely the order parameter of the bosonic cloud, $\hbar$ is the reduced Planck constant, and $m_B$ is the boson mass.  The self--interaction coupling $\tilde g=4\pi\hbar^2\tilde a_s/m_B$ depends on the $s$-wave scattering length $\tilde a_s$, while the gravitational potential $\tilde V$ is sourced by the mass density $\tilde\rho=m_B|\tilde\Psi|^2$ and its spatial average $\bar{\tilde\rho}$ defined in the spatial domain of solution $D\subset{\mathbb{R}^3}$.

\subsection{Dimensionless formulation}
\label{sec:dimensionless}

For numerical convenience and to eliminate explicit physical units, we adopt a
dimensionless formulation. Introducing a characteristic length scale $x_0$, all
dimensional quantities (denoted by tildes) are rescaled as

\[
    \tilde t = t_0\, t,\quad
    \tilde{\mathbf{x}} = x_0\, \mathbf{x},\quad
    \tilde\Psi = \sqrt{\frac{\rho_0}{m_B}}\,\Psi,\quad
    \tilde V = v_0^2\, V,\quad
    \tilde a_s = a_0\, a_s,
\]

\noindent where the characteristic scales are defined by

\begin{align}
    t_0 &= \frac{m_B x_0^{2}}{\hbar}, &
    v_0 &= \frac{\hbar}{m_B x_0},\\
    \rho_0 &= \frac{\hbar^{2}}{4\pi G m_B^{2} x_0^{4}}, &
    a_0 &= \frac{G m_B^{3} x_0^{2}}{\hbar^{2}}.
\end{align}

\noindent With these definitions, the GPP system can be written in the
dimensionless form

\begin{align}
    {\rm i}\,\partial_t\Psi
        &= \left[-\tfrac{1}{2}\nabla^2 + V + g|\Psi|^2\right]\Psi,
    \label{eq:gp-dimless}\\
    \nabla^2 V
        &= 4\pi(\rho - \bar\rho),
        \qquad \rho = |\Psi|^2,
    \label{eq:poisson-dimless}
\end{align}

\noindent where $g = 4\pi a_s$. The system is invariant under the scaling transformation

\begin{equation}
    \{t,\mathbf{x},\Psi,V,g\}
    \to
    \{\lambda^{-2} t,\;
      \lambda^{-1}\mathbf{x},\;
      \lambda^{2}\Psi,\;
      \lambda^{2}V,\;
      \lambda^{-2}g\}.
\end{equation}

\subsection{Conserved quantities and diagnostics}
\label{sec:diagnostics}

For time--dependent evolutions, we monitor the conserved particle number, which in the present normalization coincides with the total mass

\[
    M \equiv N = \int_D \rho\,d^3x.
\]

\noindent The energy is decomposed into kinetic, gravitational, and self--interaction contributions,

\begin{align}
    K &= \tfrac12\int_D |\nabla\Psi|^2\,d^3x,\\
    W &= \tfrac12\int_D \rho\,V\,d^3x,\\
    I &= \tfrac{g}{2}\int_D \rho^2\,d^3x,
\end{align}

\noindent with total energy $E=K+W+I$. Departures from equilibrium are quantified through the virial parameter

\begin{equation}    
    \eta=\frac{2K+W+3I}{|W|}.
\end{equation}

\subsection{Hydrodynamic (Madelung) formulation}
\label{sec:madelung}

A complementary hydrodynamic description of the GPP system is obtained through the Madelung transformation \citep{Shapiro2021,AlvarezGuzmanMadelung}

\begin{equation}
    \Psi=\sqrt{\rho}\,e^{{\rm i}S},
    \qquad
    \vec v=\nabla S,
\end{equation}

\noindent which maps the complex wavefunction onto a real density field $\rho$ and a velocity potential $S$. In this representation, the phase gradient $\vec v$ plays the role of a velocity field, and the dynamics can be interpreted in fluid-like terms.

Substituting this decomposition into the dimensionless Gross--Pitaevskii equation~\eqref{eq:gp-dimless} yields the continuity equation

\begin{equation}
    \partial_t\rho+\nabla\!\cdot(\rho\,\vec v)=0,
\end{equation}

\noindent together with an Euler-type equation for the velocity field,

\begin{equation}
    \partial_t\vec v+(\vec v\!\cdot\!\nabla)\vec v
        =-\nabla(V+h)+\nabla Q,
\end{equation}

\noindent where

\begin{equation}
    Q=-\frac12\frac{\nabla^2\sqrt{\rho}}{\sqrt{\rho}},
    \qquad
    h=g\rho,
\end{equation}

\noindent are the quantum potential and the specific enthalpy associated with the self-interaction term, respectively.

The quantum potential $Q$ encodes purely wave-like effects such as dispersion and interference, and has no classical fluid analogue. In particular, it is responsible for the regularization of density cusps and for the appearance of solitonic cores in BECDM halos. The enthalpy term $h=g\rho$ corresponds to a barotropic pressure

\begin{equation}
    p_{\mathrm{SI}}=\tfrac{g}{2}\rho^2,
\end{equation}

\noindent which is equivalent to a polytropic equation of state with index $n=1$ and polytropic constant $K=g/2$~\cite{AlvarezGuzman2022}.

Although the Madelung formulation provides valuable physical intuition, it becomes ill-defined at nodes of the wavefunction where $\rho=0$ and the phase is singular \cite{AlvarezGuzmanMadelung}. For this reason, all numerical evolutions in this work are carried out at the level of the wavefunction $\Psi$, while the hydrodynamic variables are used exclusively for diagnostics and physical interpretation.

%-----------------------------------> SEC: AMR
\section{AMR framework}
\label{sec:AMR}

We solve the GPP system on the computational domain

\[
D =
[x_{\min}^{(0)},x_{\max}^{(0)}]
\times
[y_{\min}^{(0)},y_{\max}^{(0)}]
\times
[z_{\min}^{(0)},z_{\max}^{(0)}],
\]

\noindent described in Cartesian coordinates. The base numerical domain is uniformly discretized with resolution

\begin{align*}
\Delta x^{(0)} &= \frac{x_{\max}^{(0)}-x_{\min}^{(0)}}{N_x^{(0)}},\\
\Delta y^{(0)} &= \frac{y_{\max}^{(0)}-y_{\min}^{(0)}}{N_y^{(0)}},\\
\Delta z^{(0)} &= \frac{z_{\max}^{(0)}-z_{\min}^{(0)}}{N_z^{(0)}}.
\end{align*}

\noindent The root grid is defined by

\begin{align*}
\mathcal{G}^{(0)} = \{(x_i,y_j,z_k)\in D \mid\;
&x_i = x_{\min}^{(0)} + i\,\Delta x^{(0)},\\
&y_j = y_{\min}^{(0)} + j\,\Delta y^{(0)},\\
&z_k = z_{\min}^{(0)} + k\,\Delta z^{(0)}\},
\end{align*}

\noindent with $i=0,\ldots,N_x^{(0)}\!-1$, $j=0,\ldots,N_y^{(0)}\!-1$, and $k=0,\ldots,N_z^{(0)}\!-1$. Time is discretized as $t^n = n\,\Delta t^{(0)}$, with

\[
\Delta t^{(0)} = C^{(0)}\,
\min\big(\Delta x^{(0)},\Delta y^{(0)},\Delta z^{(0)}\big)^{2},
\]

\noindent where $C^{(0)}$ satisfies the stability condition associated with the Laplacian operator.

The fundamental grid functions are the discrete order parameter $\Psi^n_{i,j,k} = \Psi(t^n,x_i,y_j,z_k)$, the gravitational potential $V^n_{i,j,k} = V(t^n,x_i,y_j,z_k)$, and the density $\rho^n_{i,j,k} = |\Psi^n_{i,j,k}|^2$.

The main motivation for using AMR is that a uniformly fine grid is computationally prohibitive, as both memory usage and runtime scale with the total number of grid points. In addition, Poisson solvers and multistage time integrators dominate the computational cost. Block--structured AMR addresses this limitation by refining only in regions where relevant structures develop (e.g., strong gradients or phase singularities), achieving near--uniform fine--grid accuracy at a fraction of the computational cost.

\subsection{AMR as a tree: nodes and grid functions}
\label{sec:amr-tree}

{\it Node -- tree element.}
The basic AMR unit is a tree \emph{node}. Each node stores: (i) its mesh $\mathcal{G}^{(l)}_{abc}$ and metadata (level $l$, tile position $(a,b,c)$ within its parent, local bounds and spacings, and ghost-zone layout); (ii) the \emph{grid functions} defined on that mesh (e.g., $\Psi$, $V$, $\rho$); and (iii) the $\texttt{nodes}_x\times\texttt{nodes}_y\times\texttt{nodes}_z$ array of pointers to children. Refinement allocates and initializes children, while coarsening deallocates them.

{\it Terminology: parent/child.}
A level-$(l\!-\!1)$ node is the \emph{parent} of the level-$l$ nodes obtained by tessellating its domain into $\texttt{nodes}_x\times\texttt{nodes}_y\times\texttt{nodes}_z$ equal subdomains. Each subdomain defines a \emph{child} $\mathcal{G}^{(l)}_{abc}$. The root $\mathcal{G}^{(0)}$ has no parent. Children are disjoint and, up to ghost layers, cover the parent domain. With $2{:}1$ refinement ($r=2$), parent indices $(i_p,j_p,k_p)$ align with child indices $(2i_c,2j_c,2k_c) = (i_p,j_p,k_p)$ wherever both levels are present.

{\it Child geometry.}
The triplet $(a,b,c)$ labels the child’s position within the parent tiling, with $a\in\{0,\ldots,\texttt{nodes}_x\!-1\}$ and analogously for $b$ and $c$. If the parent spans

\[
[x_{\min}^{(l-1)},x_{\max}^{(l-1)}]
\times
[y_{\min}^{(l-1)},y_{\max}^{(l-1)}]
\times
[z_{\min}^{(l-1)},z_{\max}^{(l-1)}],
\]

\noindent we define

\begin{align*}
\Delta X^{(l-1)} &= \frac{x_{\max}^{(l-1)}-x_{\min}^{(l-1)}}{\texttt{nodes}_x},\\
\Delta Y^{(l-1)} &= \frac{y_{\max}^{(l-1)}-y_{\min}^{(l-1)}}{\texttt{nodes}_y},\\
\Delta Z^{(l-1)} &= \frac{z_{\max}^{(l-1)}-z_{\min}^{(l-1)}}{\texttt{nodes}_z}.
\end{align*}

Then the child domain is

\[
x\in\big[x_{\min}^{(l-1)}+a\,\Delta X^{(l-1)},\;
          x_{\min}^{(l-1)}+(a+1)\Delta X^{(l-1)}\big],
\]

\noindent with analogous expressions for $y$ and $z$. Spatial spacings refine with ratio $r=2$: $\Delta x^{(l)} = \Delta x^{(l-1)}/2$, and similarly for $y$ and $z$.

{\it Node creation -- refinement.}
Refinement is triggered when local indicators exceed a prescribed threshold. In this work, refinement is driven by physically motivated criteria based on the solution variables. The tagged parent at level $(l\!-\!1)$ is partitioned into $\texttt{nodes}_x\times\texttt{nodes}_y\times\texttt{nodes}_z$ children $\mathcal{G}^{(l)}_{abc}$. Time integration is subcycled with $\Delta t^{(l)} = \tfrac{1}{2}\,\Delta t^{(l-1)}$.

{\it Prolongation (parent$\to$child).} 
Let $f$ be any grid function defined on a level-$(l-1)$ parent block. For $2{:}1$ refinement, each child index $(i_c,j_c,k_c)$ can be written as an aligned (coarse) index plus a parity,

\[
i_c = 2i_p + \varepsilon_i,\quad
j_c = 2j_p + \varepsilon_j,\quad
k_c = 2k_p + \varepsilon_k,\quad
\]

\noindent with $\varepsilon_i,\varepsilon_j,\varepsilon_k\in\{0,1\}$, where $\varepsilon=0$ denotes an aligned point and $\varepsilon=1$ an
intermediate point between two parent nodes.

Prolongation is performed by a tensor-product stencil.  In each coordinate direction we define the 1D index set

\[
S(\varepsilon)=
\begin{cases}
\{0\}, & \varepsilon=0,\\[2pt]
\{-2,-1,0,1,2,3\}, & \varepsilon=1,
\end{cases}
\]

\noindent and the corresponding selector

\[
\theta_\alpha(\varepsilon)=
\begin{cases}
1, & \varepsilon=0,\ \alpha=0,\\
0, & \varepsilon=0,\ \alpha\neq 0,\\
w_\alpha, & \varepsilon=1,
\end{cases}
\]

\noindent with sixth-point interpolation weights

\begin{align*}
w_{-2}=w_{3}=\frac{3}{256},\\
w_{-1}=w_{2}=-\frac{25}{256},\\
w_{0}=w_{1}=\frac{150}{256}.
\end{align*}

\noindent With these definitions, the prolongated value at the child node $(i_c,j_c,k_c)$ is

\begin{align}
f^{(l)}_{i_c,j_c,k_c}
& =
\sum_{\alpha\in S(\varepsilon_i)}
\sum_{\beta \in S(\varepsilon_j)}
\sum_{\gamma\in S(\varepsilon_k)} \nonumber \\
& \quad \theta_\alpha(\varepsilon_i)\,
\theta_\beta(\varepsilon_j)\,
\theta_\gamma(\varepsilon_k)\,
f^{(l-1)}_{\,i_p+\alpha,\;j_p+\beta,\;k_p+\gamma}.
\label{eq:prolongation-6pt}
\end{align}

In particular, if $(\varepsilon_i,\varepsilon_j,\varepsilon_k)=(0,0,0)$ the stencil reduces to a direct copy,

\[
f^{(l)}_{i_c,j_c,k_c}=f^{(l-1)}_{i_p,j_p,k_p}.
\]

\noindent Child ghost zones are filled by applying the same tensor-product stencil, restricted to the portion of the parent block that overlaps the child (including the required parent ghost layers).

For intermediate times $t\in[t^n,t^{n+1}]$ we use linear interpolation on the parent grid,

\[
f^{(l-1)}(t,\mathbf{x})
\simeq
(1-\theta)\,f^{(l-1)}(t^n,\mathbf{x})
+
\theta\,f^{(l-1)}(t^{n+1},\mathbf{x}),
\]

\noindent with $\theta = \frac{t-t^n}{\Delta t^{(l-1)}}$.

{\it Injection child$\to$parent.} Pointwise restriction is applied to update the parent nodes covered by the child grid,

\[
\Psi^{(l-1)}_{i_p,j_p,k_p}
    \;\leftarrow\;
    \Psi^{(l)}_{i_c,j_c,k_c},
\]

\noindent thereby restoring inter level consistency before the next parent-level update.

\subsection{Time integration}
\label{sec:integration}

The AMR hierarchy is evolved in time using Berger--Oliger subcycling \cite{berger1984amr}, combined with an explicit fourth--order Runge--Kutta (RK4) scheme. At each refinement level~$l$, the wavefunction is advanced from $t^n$ to $t^{n+1}=t^n+\Delta t^{(l)}$ according to

\begin{equation}
    \Psi^{n+1}
    =
    \Psi^{n}
    +
    \frac{\Delta t^{(l)}}{6}
    \left(
        k_1 + 2k_2 + 2k_3 + k_4
    \right),
\end{equation}

\noindent where the intermediate slopes are given by

\begin{align}
    k_1 &= \mathcal{R}(\Psi^{n}), \\
    k_2 &= \mathcal{R}\!\left(\Psi^{n} + \tfrac{1}{2}\Delta t^{(l)} k_1\right), \\
    k_3 &= \mathcal{R}\!\left(\Psi^{n} + \tfrac{1}{2}\Delta t^{(l)} k_2\right), \\
    k_4 &= \mathcal{R}\!\left(\Psi^{n} + \Delta t^{(l)} k_3\right),
\end{align}

\noindent and the right--hand-side operator is defined as

\begin{equation}
    \mathcal{R}(\Psi)
    :=
    -{\rm i}\left(
        -\tfrac{1}{2}\nabla^2
        + V
        + g|\Psi|^2
    \right)\Psi,
\end{equation}

\noindent where the Laplacian is discretized using fourth--order finite differences.

At each RK4 substep, the gravitational potential $V$ is updated by solving Poisson equation~(\ref{eq:poisson-dimless}) using the same fourth--order finite--difference stencil. The numerical strategy adopted to solve the Poisson equation depends on the AMR level.

{\it Root level ($l=0$).}
On the root grid, periodic boundary conditions are imposed and the Poisson equation is solved spectrally using a Fast Fourier Transform (FFT) \cite{rodriguezlara2024,periodicas}. In Fourier space, the discrete fourth--order Laplacian is diagonal, so each mode of the gravitational potential can be obtained algebraically. The solution is then transformed back to real space by means of an inverse FFT.

{\it Refined levels ($l>0$).}
On refined grids, the Poisson equation is solved locally using a multigrid V-cycle \cite{GuzmanAlvarezGonzalez2021,AlvarezGuzman2022}, where successive over--relaxation (SOR) is employed as the smoothing operator for both the gravitational potential and the associated error equation. The same fourth--order finite--difference discretization is used across all levels. Dirichlet boundary conditions are supplied by the parent level through spatial prolongation and linear interpolation in time. 

\subsection{Frequency filtering}
\label{sec:filtering}

Consider, without loss of generality, the spatial direction~$x$. At refinement level~$l$, the Nyquist frequency associated with the grid spacing $\Delta x^{(l)}$ is given by~\cite{rodriguezlara2024}
\begin{equation}
    f_c^{(l)} = \frac{1}{2\,\Delta x^{(l)}},
\end{equation}
while for its parent level $(l-1)$ one has
\begin{equation}
    f_c^{(l-1)} = \frac{1}{2\,\Delta x^{(l-1)}}.
\end{equation}

For a standard $2{:}1$ refinement ratio, $\Delta x^{(l)} = \Delta x^{(l-1)}/2$, which implies
\begin{equation}
    f_c^{(l)} = 2\,f_c^{(l-1)}.
\end{equation}

This difference implies that child grids can represent spatial frequencies up to twice those admissible on their parent grid. Consequently, when fine-level data are injected into the coarser level, Fourier modes with $f > f_c^{(l-1)}$ cannot be properly represented and are instead aliased into lower frequencies. This aliasing introduces spurious large-scale power and may degrade both the accuracy and stability of the numerical scheme, particularly in the presence of sharp gradients, interference patterns, or highly oscillatory configurations.

To mitigate this effect, we introduce an explicit dissipation mechanism by modifying the right-hand side of the evolution equations as~\cite{refId0}
\begin{equation}
    \mathcal{R}_{\mathrm{KO}}(\Psi)
    =
    \mathcal{R}(\Psi)
    -
    \frac{(-1)^{N}}{\Delta t}
    \left(\frac{h}{\pi \varepsilon_{\mathrm{KO}}}\right)^{2N}
    \nabla^{2N}\Psi,
\end{equation}
where $\varepsilon_{\mathrm{KO}} \sim \mathcal{O}(1)$ is a dimensionless coefficient, and for simplicity we assume $\Delta x = \Delta y = \Delta z = h$. The additional term corresponds to a Kreiss--Oliger dissipation operator of order $2N$~\cite{KreissOliger1973}.

The action of this operator is most naturally understood in Fourier space. Consider the model problem
\[
\partial_t \Psi =
-
\frac{(-1)^{N}}{\Delta t}
\left(\frac{h}{\pi \varepsilon_{\mathrm{KO}}}\right)^{2N}
\nabla^{2N}\Psi.
\]
Each Fourier mode evolves independently according to
\begin{equation}
    \frac{d\hat{\Psi}}{dt}
    =
    -\frac{1}{\Delta t}
    \left(\frac{k}{\pi \varepsilon_{\mathrm{KO}}}\right)^{2N}
    \hat{\Psi},
\end{equation}
whose solution over a single time step $\Delta t$ is
\begin{equation}
    \hat{\Psi}^{\,n+1}
    =
    \exp\!\left[
    -\left(\frac{k}{\varepsilon_{\mathrm{KO}}\,k_c}\right)^{2N}
    \right]
    \hat{\Psi}^{\,n},
    \label{eq:filter}
\end{equation}
where $k_c = 2\pi f_c = \pi / h$ is the Nyquist wavenumber. Therefore, the dissipation term acts as a spectral low-pass filter: long-wavelength modes ($k \ll k_c$) are essentially unaffected, while modes near the grid scale ($k \sim k_c$) are exponentially damped.

This spectral interpretation provides a natural criterion for relating the strength of the dissipation to the local grid resolution. In particular, increasing $N$ makes the filter more selective, concentrating the dissipative action on the highest resolvable frequencies.

In this work, we select the order of the Kreiss--Oliger operator according to the condition
\begin{equation}
    2N - 2 > p,
\end{equation}
where $p$ denotes the order of accuracy of the underlying finite--difference scheme. In our case, $p = 4$, so we choose $N = 4$, corresponding to an eighth--order Kreiss--Oliger operator.

This choice ensures that the dissipation term is of order at least $p+1$, independently of the formal order of the differential operator $\nabla^{2N}$. As a result, the artificial dissipation converges to zero faster than the truncation error of the numerical scheme in the limit $h \to 0$, thereby preserving consistency. In particular,
\begin{equation}
    \mathcal{R}_{\mathrm{KO}}(\Psi) \to \mathcal{R}(\Psi)
    \quad \text{as} \quad h \to 0.
\end{equation}

The dissipation is applied at all refinement levels to maintain consistency of the right-hand side of the evolution equations across the AMR hierarchy. This ensures uniform spectral filtering and improves the robustness of parent--child data transfer.

\subsection{Parallel implementation}

The code is implemented in Fortran and parallelized using the Message Passing Interface (MPI) for distributed-memory architectures. The computational domain is decomposed across processes using a two-dimensional pencil decomposition aligned with the $z$-axis. This strategy is particularly well suited for the implementation of distributed three-dimensional FFTs, as required by the Poisson solver at the root level. The approach follows standard parallel FFT techniques based on global data redistributions, enabling efficient transpositions between different data layouts (see, e.g., \cite{DALCIN2019137}).

Each MPI process evolves a local subdomain of the global grid, augmented with ghost zones that store boundary data from neighboring subdomains. The number of ghost cells is chosen to be consistent with the fourth-order finite-difference stencil employed in the spatial discretization.

Simulation data are written in both ASCII and HDF5 formats. The HDF5 output is organized hierarchically following the AMR tree structure, enabling efficient storage and post-processing of multi-level data. This format is compatible with standard visualization tools such as \texttt{VisIt}.

To characterize the parallel performance of the implementation, we performed strong-scaling tests using the equilibrium benchmark for fixed problem sizes of $64^3$ and $128^3$. To reduce dependence on hardware characteristics, execution times are normalized using the wall-clock time measured for the $64^3$ unigrid simulation evolved with a single MPI process.

\begin{table}
\centering
\begin{tabular}{ccccc}
\hline
Resolution &
MPI &
Runtime &
Speedup &
Efficiency
\\
\hline

$64^3$ & 1 & 1.00 & 1.00 & 1.00 \\

$64^3$ & 2 & 0.512 & 1.95 & 0.98 \\

$64^3$ & 4 & 0.253 & 3.95 & 0.99 \\

$64^3$ & 8 & 0.131 & 7.62 & 0.95 \\

$64^3$ & 16 & 0.070 & 14.34 & 0.90 \\

$64^3$ & 32 & 0.042 & 23.72 & 0.74 \\

\hline

$128^3$ & 1 & 8.71 & 1.00 & 1.00 \\

$128^3$ & 2 & 4.31 & 2.02 & 1.01 \\

$128^3$ & 4 & 2.11 & 4.13 & 1.03 \\

$128^3$ & 8 & 1.11 & 7.81 & 0.98 \\

$128^3$ & 16 & 0.57 & 15.23 & 0.95 \\

$128^3$ & 32 & 0.31 & 27.96 & 0.87 \\

\hline
\end{tabular}
\caption{
Strong-scaling results for fixed problem sizes. Execution times are normalized to the wall-clock time of the $64^3$ unigrid run using one MPI process. The speedup $S_p$ is defined as $S_p=T_1/T_p$, where $T_1$ and $T_p$ denote the execution times obtained using one (with $64^3$ grid resolution) and $p$ MPI processes, respectively. The parallel efficiency $E_p$ is defined as $E_p=S_p/p$, where $p$ is the number of MPI processes.}
\label{tab:strong_scaling}
\end{table}

Table~\ref{tab:strong_scaling} shows that the parallel implementation exhibits near-ideal strong scaling over the range of processor counts explored in this work.

For the $64^3$ benchmark, efficiencies remain above $90\%$ up to 16 MPI processes and decrease to approximately $74\%$ at 32 processes, indicating the onset of communication overhead.

The larger $128^3$ problem exhibits improved scalability, maintaining efficiencies above $95\%$ up to 16 processes and approximately $87\%$ at 32 processes.

These results indicate that the pencil decomposition and communication strategy remain effective across the explored processor range and support efficient large-scale distributed simulations.

%-------------------> SEC: TESTS
\section{Validation and test problems}
\label{sec:validation}

We validate the AMR solver for the GPP system, Eqs.~(\ref{eq:gp-dimless})--(\ref{eq:poisson-dimless}), through five test problems designed to probe its performance in fully nonlinear regimes. These tests assess the accuracy of the time integration scheme, the consistency of interlevel synchronization, and the robustness of the Poisson solver under periodic boundary conditions. Unless otherwise stated, all simulations employ the dimensionless formulation introduced in Sec.~\ref{sec:model}. Global diagnostics include mass conservation, energy conservation, and the virial parameter.

Specifically, we consider the following tests:
\begin{enumerate}
    \item Advection of a boosted solitonic core.
    \item Advection of a boosted line vortex.
    \item Merger of a binary system of solitonic cores.
    \item Merger of a binary system of line vortices.
    \item Gravitational condensation of a random bosonic cloud.
\end{enumerate}

The first four tests are performed in the nonlinear regime with repulsive self--interaction, $g=1$. This choice allows us to probe nonlinear effects while avoiding additional dynamical instabilities associated with attractive interactions ($g<0$). The fifth test employs a different physical setup, corresponding to a homogeneous and isotropic bosonic cloud with random phases in momentum space and vanishing self--interaction ($g=0$). The numerical and physical parameters of this configuration are described in Sec.~\ref{sec:condensation}.

Each test is performed using three configurations: a uniform grid (unigrid), AMR with one refinement level, and AMR with two refinement levels, allowing for a direct comparison between uniform and adaptively refined evolutions. In all cases, the computational domain is a periodic cube of side length $L=40$, discretized on the root level with $N=64$ grid points per spatial direction. Time integration follows a Berger--Oliger subcycling strategy with Courant factor

\begin{equation}
C^{(l)} = \frac{0.256}{2^{\,l}}, \qquad l=0,1,2.
\end{equation}

Adaptive refinement is driven by a physically motivated criterion based on the magnitude of the gravitational potential. In the GPP system, the potential $V$ encodes the global distribution of mass through Poisson equation, and therefore provides a nonlocal measure of dynamically relevant regions.

In particular, regions of large $|V|$ correspond to gravitationally bound structures such as solitonic cores, vortex cores, and zones of strong wave interference, where high spatial resolution is required to accurately capture both the density and phase structure of the wavefunction.

A block $\mathcal{G}^{(l)}_{abc}$ is flagged for refinement whenever

\begin{equation}
\frac{|V|}
     {\max_{\mathcal{G}^{(l)}} |V|}
> \varepsilon,
\qquad \varepsilon = 0.75,
\end{equation}

\noindent which concentrates resolution in regions of strong gravitational potential, such as solitonic cores, vortex cores, and interference patterns.

To control high-frequency numerical noise and mitigate aliasing effects near refinement boundaries, we apply an eighth-order Kreiss--Oliger operator, equivalent to a Gaussian low-pass filter. In all validation tests, the filter strength is chosen within the range $\epsilon_{\mathrm{KO}} \sim \mathcal{O}(1)$, providing effective damping of near--Nyquist modes without affecting the resolved physical dynamics.

The first three tests are evolved up to $t_{\max}=100$ code units, which is sufficient to capture the characteristic oscillation, advection, and virialization timescales of the system. The fourth test, involving the merger of vortex binaries, is evolved up to $t_{\max}=200$ to resolve the extended nonlinear interaction and subsequent relaxation. The fifth test, corresponding to gravitational Bose--Einstein condensation, employs a different computational setup and integration time, as described in Sec.~\ref{sec:condensation}.

%------------------------------------------------------------

\subsection{Boosted soliton: periodic advection}
\label{sec:boosted-soliton}

{\it Purpose.}
This test assesses the ability of the AMR GPP solver to advect a localized, self--gravitating equilibrium configuration across a periodic domain while preserving its internal structure and global conserved quantities. It provides a direct comparison between uniform and adaptive evolutions in a fully nonlinear regime, and evaluates whether AMR reproduces the same physical dynamics without introducing spurious dissipation or secular drift.

\medskip

{\it Initial conditions.}
The initial wavefunction corresponds to a stationary solitonic core constructed from the equilibrium profiles summarized in Appendix~\ref{app:stationary-cores}. A uniform Galilean boost is applied along the $x$ direction,
\begin{equation}
\Psi(0,\mathbf{x})
=
\sqrt{\rho_{\mathrm{core}}(r)}\,e^{i v_0 x}.
\end{equation}

\medskip

\begin{figure}
    \centering
    \includegraphics[width=8cm]{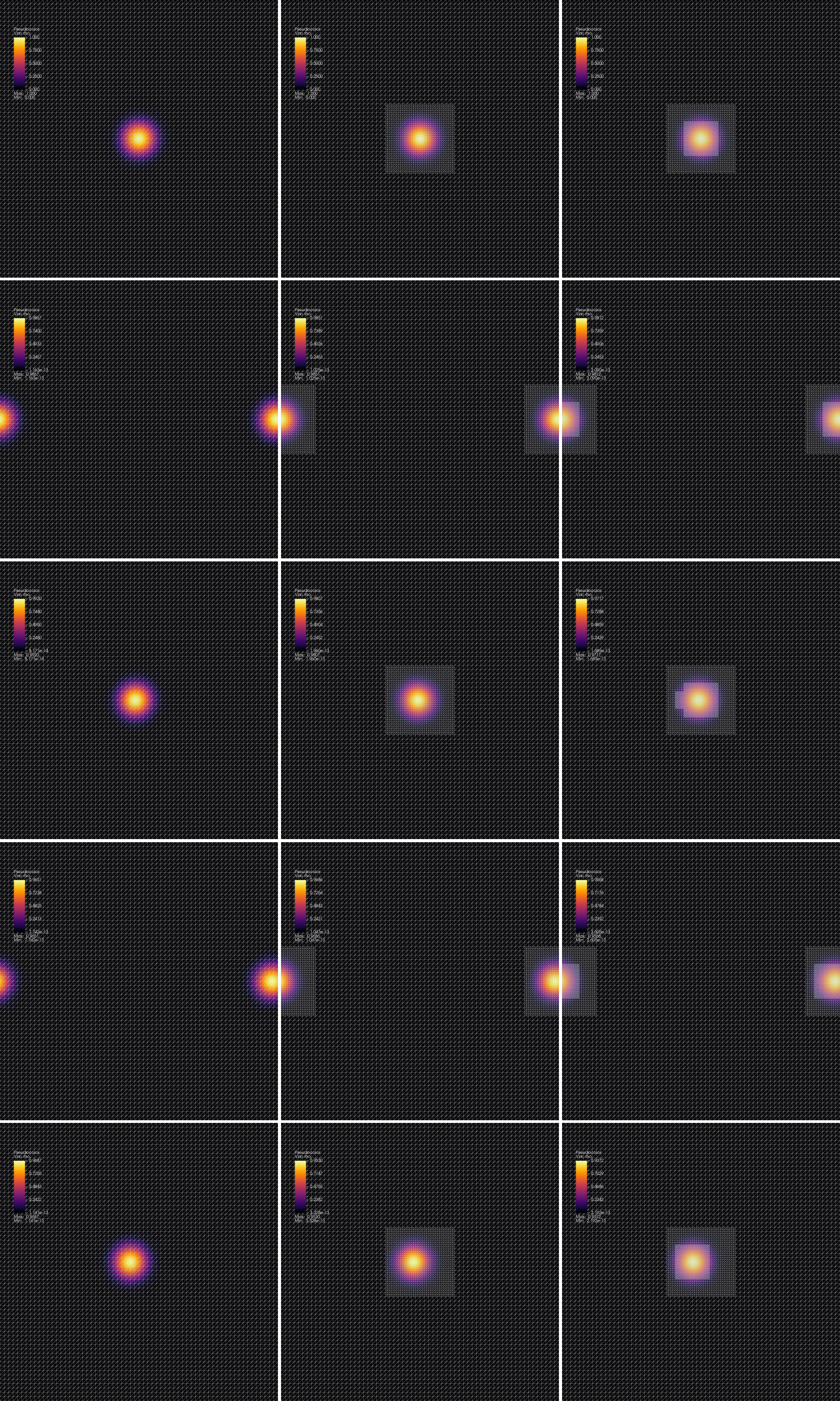}
    \caption{Advection of a boosted solitonic core under periodic boundary conditions for $g=1$. Density slices on the midplane $z=0$ are shown together with the computational mesh. Columns correspond to the unigrid evolution (left), AMR with one refinement level (center), and AMR with two refinement levels (right). Rows show snapshots at $t=0$, $20$, $40$, $60$, and $80$ (top to bottom), illustrating coherent advection and close agreement across all configurations.}
    \label{fig:boosted_soliton_evolution}
\end{figure}

{\it Results.}
Figure~\ref{fig:boosted_soliton_evolution} shows the time evolution of the boosted soliton for three configurations: a uniform grid (first column), AMR with one refinement level (second column), and AMR with two refinement levels (third column). Rows display snapshots at $t=0$, $20$, $40$, $60$, and $80$.

The solitonic core is advected coherently across the periodic domain without noticeable deformation, numerical diffusion, or grid imprinting. The AMR hierarchy remains well localized around the soliton and adapts smoothly as it traverses the domain. The morphology and trajectory are indistinguishable across all configurations, indicating that the adaptive scheme faithfully reproduces the unigrid solution.

\medskip

\begin{figure}
    \centering
    \includegraphics[width=8cm]{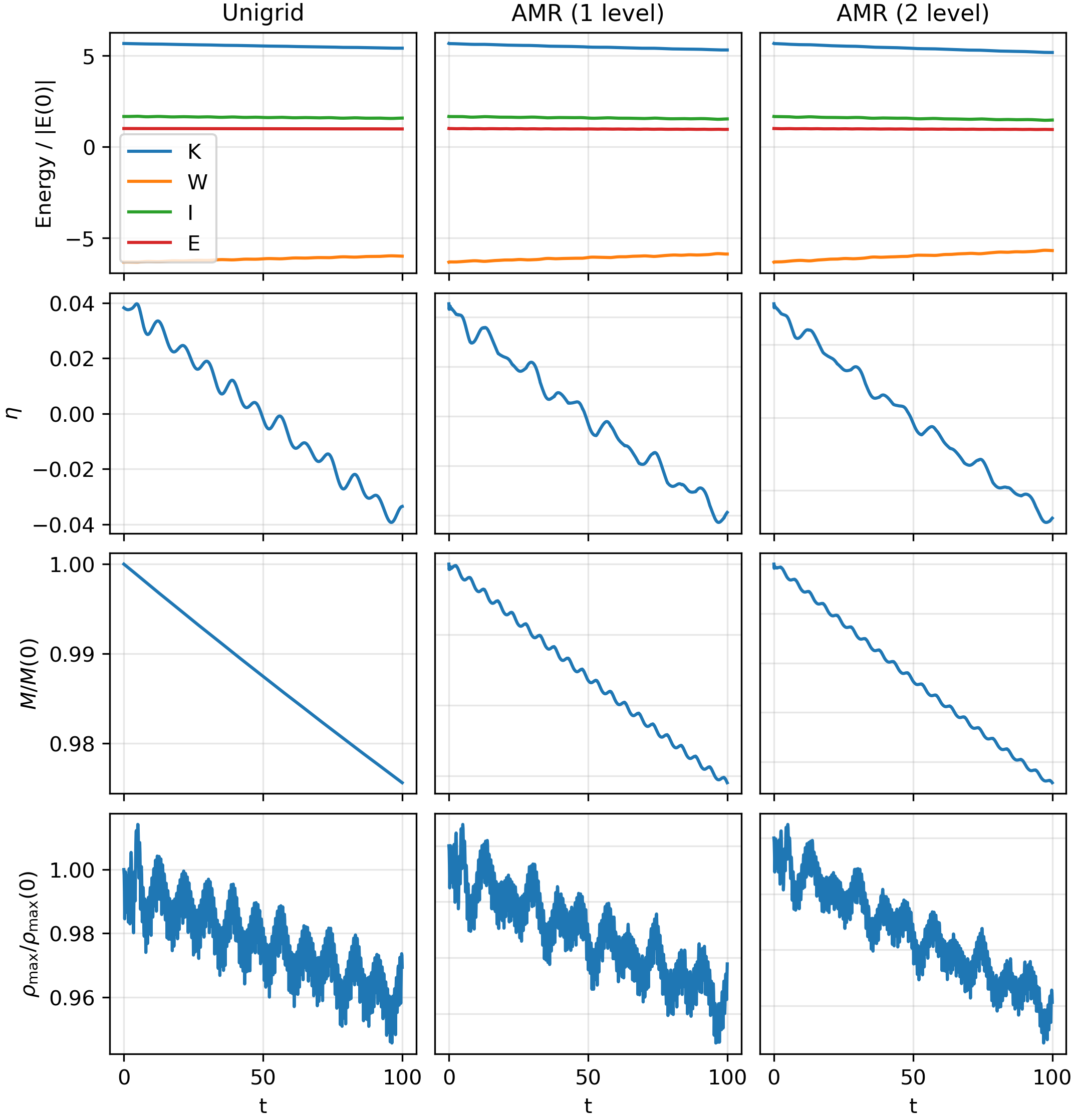}
    \caption{Global diagnostics for the boosted soliton test with $g=1$. Columns correspond to the unigrid evolution (left), AMR with one refinement level (center), and AMR with two refinement levels (right). Rows show, from top to bottom: energy components normalized to $|E(0)|$, virial parameter $\eta$, total mass $M/M(0)$, and maximum density $\rho_{\max}/\rho_{\max}(0)$ as functions of time. The agreement across configurations demonstrates excellent conservation properties and long-term stability of the AMR scheme.}
    \label{fig:boosted_soliton_diagnostics}
\end{figure}

{\it Diagnostics.}
Global diagnostics are shown in Fig.~\ref{fig:boosted_soliton_diagnostics}, with columns corresponding to the unigrid evolution, AMR with one refinement level, and AMR with two refinement levels.

The first row displays the kinetic, gravitational, self--interaction, and total energies, normalized by $|E(0)|$. All components remain nearly constant, with small bounded oscillations associated with intrinsic breathing modes of the stationary soliton.

The second row shows the virial parameter $\eta$, which oscillates around zero, indicating that the system remains close to virial equilibrium throughout the evolution. The third row presents the total mass $M/M(0)$, which is well conserved, with only negligible deviations. The fourth row shows the maximum density $\rho_{\max}(t)/\rho_{\max}(0)$, exhibiting small oscillations consistent with the same breathing dynamics.

The close quantitative agreement across all configurations demonstrates that adaptive refinement preserves both the physical dynamics and the global conservation properties of the system during nonlinear soliton advection.

Additional long-term evolutions of this benchmark were performed up to $t=1000$ for different values of the Kreiss--Oliger parameter $\epsilon_{\rm KO}$. The same test was also used to quantify the computational overhead of adaptive refinement in terms of normalized runtime and peak memory usage. The corresponding mass-conservation and performance analyses are presented in Appendix~\ref{app:longterm_mass}.

\subsection{Boosted stationary line vortex}
\label{sec:boosted-line-vortex}

{\it Purpose.}
This test assesses the ability of the AMR GPP solver to advect a stationary topological defect across a periodic domain while preserving its internal structure and phase singularity. In contrast to the boosted soliton test, which probes the transport of a localized equilibrium, this benchmark focuses on the evolution of a straight quantized vortex line with nontrivial phase winding. It provides a stringent test of time integration, periodic boundary conditions, and adaptive mesh refinement in the presence of a topological defect.

\medskip

{\it Initial conditions.}
The initial configuration corresponds to a stationary straight vortex line aligned with the $z$ axis, constructed following Appendix~\ref{app:line-vortices}. The wavefunction is initialized using the axisymmetric ansatz
\begin{equation}
\Psi(0,r_\perp,\varphi,z)
=
\psi(r_\perp,z)\,e^{i m \varphi},
\end{equation}
with winding number $m=1$. The real amplitude $\psi(r_\perp,z)$ is obtained as a stationary solution of the GPP system via imaginary--time evolution, with equatorial symmetry $\partial_z\psi(r_\perp,0)=0$ and the vortex core condition $\psi(0,z)=0$. The total mass is normalized to $M \approx 25$.

A uniform Galilean boost is applied along the $x$ direction,
\begin{equation}
\Psi(0,\mathbf{x})
\;\longrightarrow\;
\Psi(0,\mathbf{x})\,e^{i v_{x0} x},
\end{equation}
with $v_{x0}=1$, matching the boosted soliton test.

\medskip

\begin{figure}
    \centering
    \includegraphics[width=8cm]{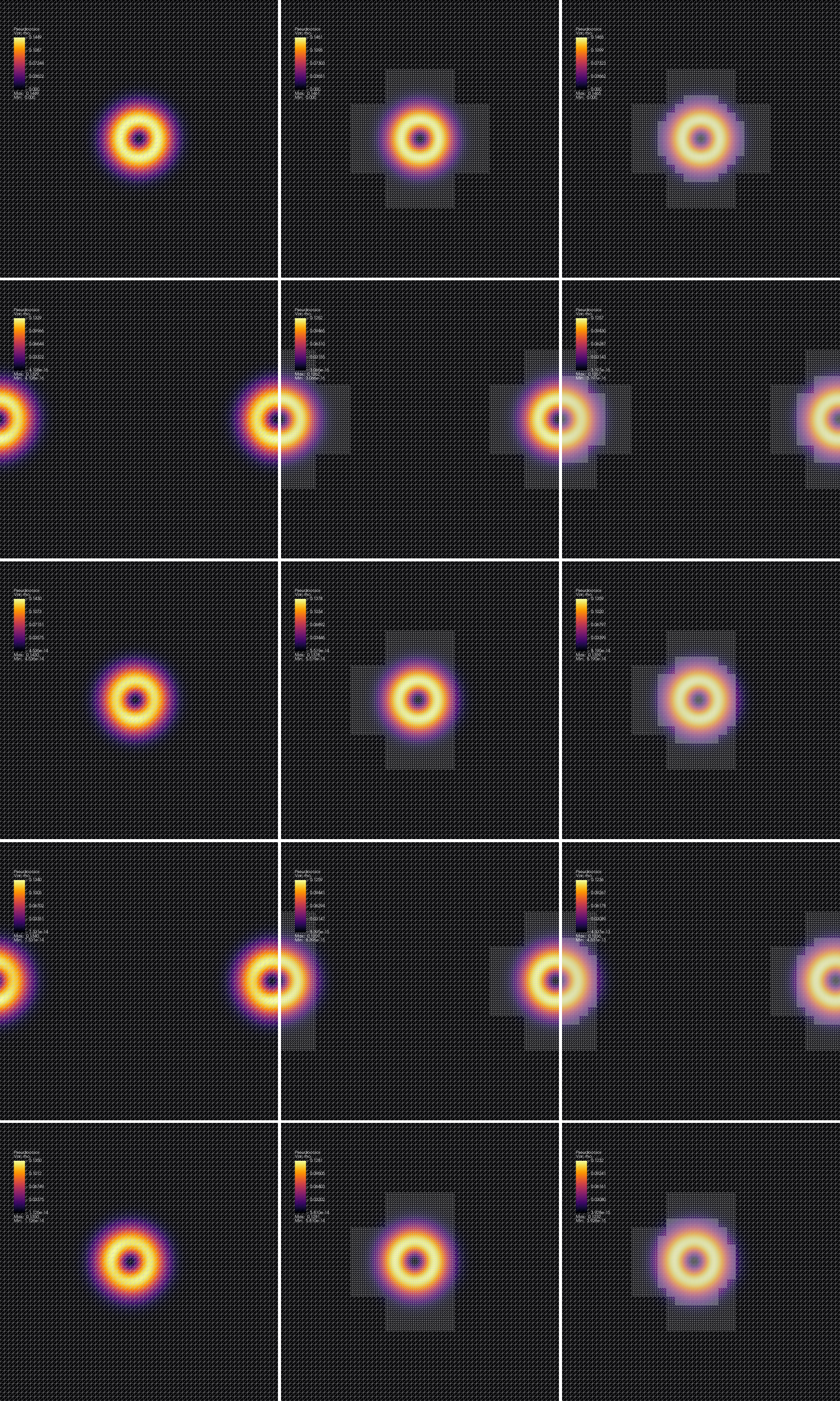}
    \caption{Advection of a boosted stationary line vortex under periodic boundary conditions for $g=1$. Density slices on the midplane $z=0$ are shown together with the computational mesh. Columns correspond to the unigrid evolution (left), AMR with one refinement level (center), and AMR with two refinement levels (right). Rows show snapshots at $t=0$, $20$, $40$, $60$, and $80$ (top to bottom), demonstrating stable transport and preservation of the vortex core structure.}
    \label{fig:vortex_evolution}
\end{figure}

{\it Results.}
Figure~\ref{fig:vortex_evolution} shows the evolution of the boosted vortex for three configurations: unigrid (left), AMR with one refinement level (center), and AMR with two refinement levels (right). Rows display snapshots at $t=0$, $20$, $40$, $60$, and $80$.

The vortex is advected coherently across the domain without distortion of the annular density profile or loss of the central phase singularity. No drift, unwinding, or fragmentation is observed. The AMR hierarchy remains well localized and tracks both the vortex core and its associated gravitational potential without introducing refinement artifacts. The evolution is indistinguishable across all configurations, indicating that AMR accurately reproduces the unigrid solution even in the presence of a topological defect.

\medskip

\begin{figure}
    \centering
    \includegraphics[width=8cm]{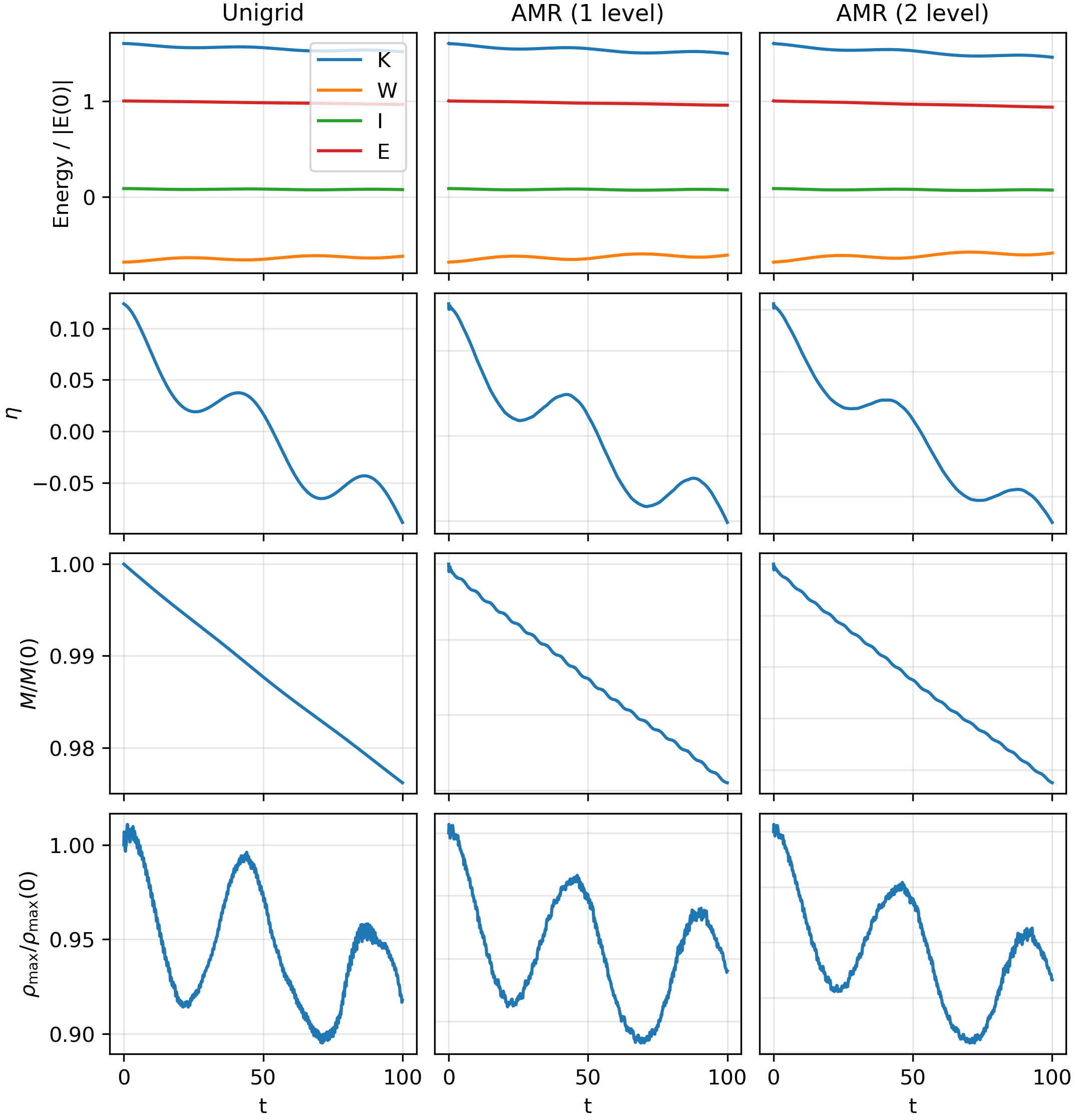}
    \caption{Global diagnostics for the boosted stationary line vortex test with $g=1$. Columns correspond to the unigrid evolution (left), AMR with one refinement level (center), and AMR with two refinement levels (right). Rows show, from top to bottom: energy components normalized to $|E(0)|$, virial parameter $\eta$, total mass $M/M(0)$, and maximum density $\rho_{\max}/\rho_{\max}(0)$ as functions of time. The agreement across configurations demonstrates stable advection and long-term preservation of the vortex structure.}
    \label{fig:vortex_diagnostics}
\end{figure}

{\it Diagnostics.}
Global diagnostics are shown in Fig.~\ref{fig:vortex_diagnostics}, with columns corresponding to the unigrid evolution, AMR with one refinement level, and AMR with two refinement levels.

The energy components remain well controlled, exhibiting small bounded oscillations associated with intrinsic breathing modes of the stationary vortex configuration. The virial parameter $\eta$ oscillates around zero, indicating that the system remains close to virial equilibrium. The total mass $M/M(0)$ is well conserved, with a small and controlled decrease attributable to Kreiss--Oliger dissipation. The maximum density $\rho_{\max}/\rho_{\max}(0)$ shows periodic oscillations without any secular drift.

The close agreement across all configurations confirms that the AMR implementation preserves both the topological structure and the global conservation properties during vortex advection.

This test demonstrates that the AMR GPP solver accurately transports stationary topological defects while preserving their phase singularity, internal structure, and global conservation properties across all refinement levels.

\subsection{Merger of two solitonic cores}
\label{sec:core-collision}

{\it Purpose.}
This test probes the capability of the AMR GPP solver to capture strongly nonlinear, non-axisymmetric dynamics involving self--gravity, wave interference, and angular momentum transport. Unlike the previous advection tests, this benchmark explores a genuinely dynamical regime in which two initially isolated solitonic cores interact and merge into a single bound configuration. It provides a stringent assessment of interlevel synchronization, prolongation and restriction operators, and the robustness of the Poisson coupling during violent interactions.

\medskip

{\it Initial conditions.}
The initial configuration consists of two identical stationary solitonic cores constructed from the equilibrium profiles described in Appendix~\ref{app:stationary-cores}. The cores are placed symmetrically in the $(x,y)$ plane at
\[
\boldsymbol{x}_1 = (-7,-7,0), \qquad
\boldsymbol{x}_2 = (+7,+7,0),
\]
and assigned equal and opposite velocities along the $x$ direction,
\[
\boldsymbol{v}_1 = (+0.2,0,0), \qquad
\boldsymbol{v}_2 = (-0.2,0,0),
\]
resulting in a finite impact parameter and nonzero orbital angular momentum about the $z$ axis.

The total wavefunction is initialized as the coherent superposition
\begin{equation}
\Psi(0,\boldsymbol{x}) =
\sqrt{\rho_{\mathrm{core}}(|\boldsymbol{x}-\boldsymbol{x}_1|)}\,
e^{i\,\boldsymbol{v}_1\cdot\boldsymbol{x}}
+
\sqrt{\rho_{\mathrm{core}}(|\boldsymbol{x}-\boldsymbol{x}_2|)}\,
e^{i\,\boldsymbol{v}_2\cdot\boldsymbol{x}}.
\end{equation}

\medskip

\begin{figure}
    \centering
    \includegraphics[width=8cm]{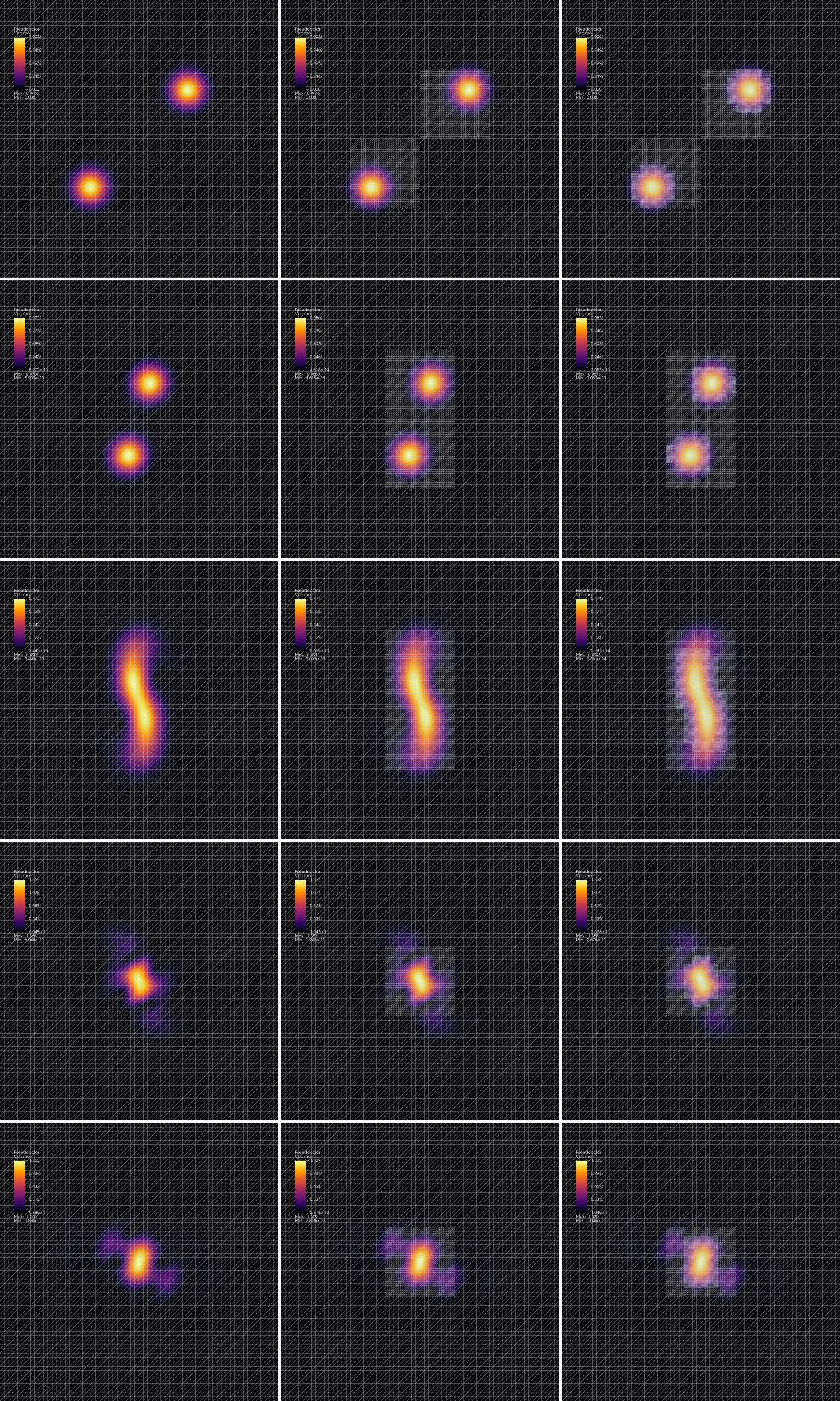}
    \caption{Merger of two solitonic cores under periodic boundary conditions for $g=1$. Density slices on the midplane $z=0$ are shown on a logarithmic scale to highlight interference patterns. Columns correspond to the unigrid evolution (left), AMR with one refinement level (center), and AMR with two refinement levels (right). Rows show snapshots at $t=0$, $20$, $40$, $60$, and $80$ (top to bottom), illustrating orbital motion, strong interference, and post--merger relaxation.}
    \label{fig:binary_evolution}
\end{figure}

{\it Results.}
Figure~\ref{fig:binary_evolution} shows the evolution for three configurations: unigrid (left), AMR with one refinement level (center), and AMR with two refinement levels (right), with snapshots at $t=0$, $20$, $40$, $60$, and $90$.

Initially, the cores approach each other and generate a rotating interference pattern. As the interaction strengthens, the cores deform and partially overlap, producing strong density modulations and phase interference. At later times, the system relaxes toward a single bound remnant surrounded by low-amplitude wave interference.

The AMR hierarchy dynamically tracks the regions of strong self--gravity and interference, remaining localized while resolving both the core structure and extended wave patterns. The morphology, phase structure, and merger outcome are consistent across all configurations, with no evidence of grid imprinting or artificial symmetry breaking. This demonstrates that AMR faithfully reproduces the unigrid solution even in strongly nonlinear, multi-scale interactions.

\medskip

\begin{figure}
    \centering
    \includegraphics[width=8cm]{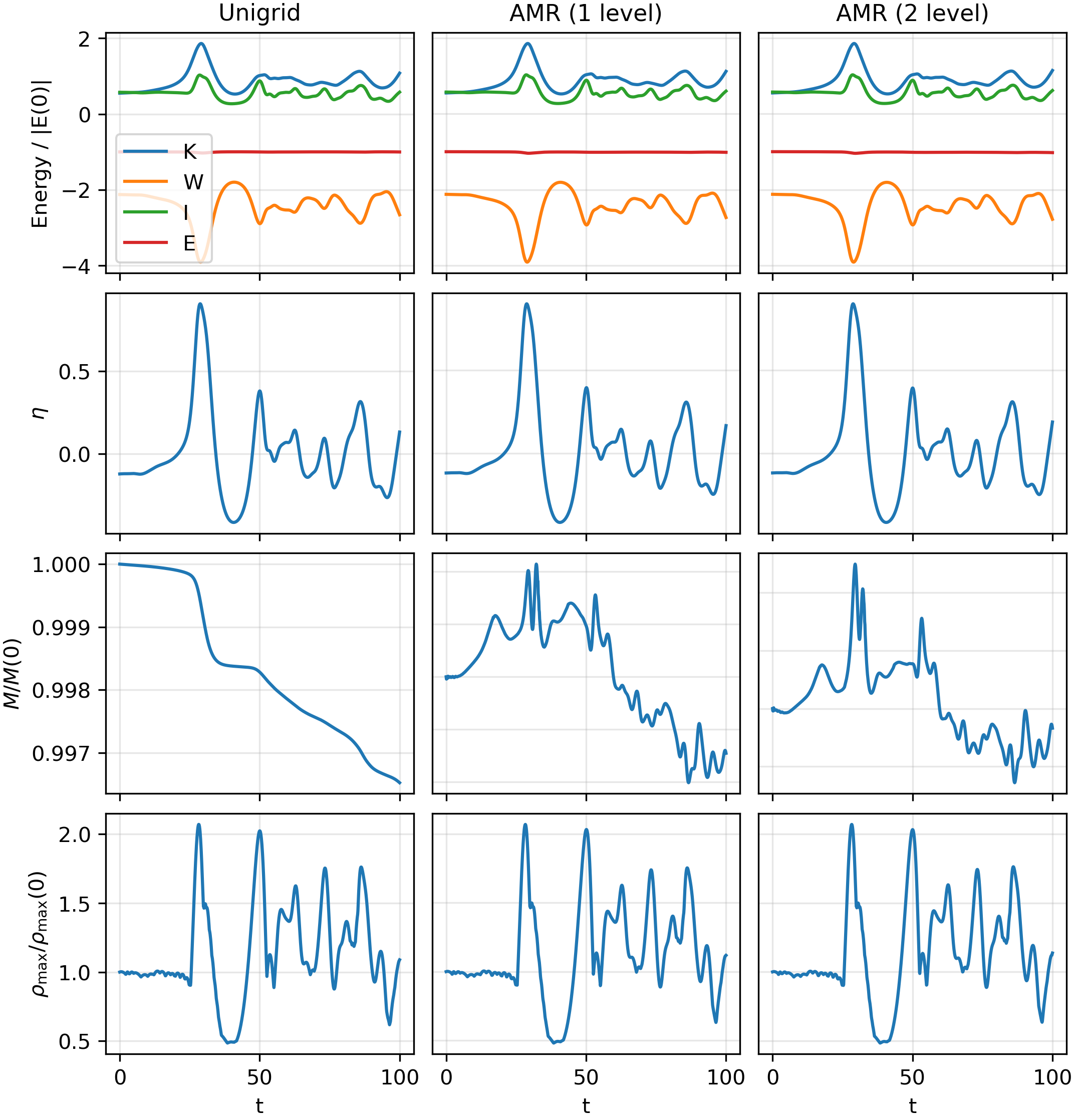}
    \caption{Global diagnostics for the merger of two solitonic cores with $g=1$. Columns correspond to the unigrid evolution (left), AMR with one refinement level (center), and AMR with two refinement levels (right). Rows show, from top to bottom: energy components normalized to $|E(0)|$, virial parameter $\eta$, total mass $M/M(0)$, and maximum density $\rho_{\max}/\rho_{\max}(0)$ as functions of time. The diagnostics capture the nonlinear interaction and relaxation toward a virialized remnant, with excellent agreement across all configurations.}
    \label{fig:binary_evolution_physical_quantities}
\end{figure}

{\it Diagnostics.}
Global diagnostics are shown in Fig.~\ref{fig:binary_evolution_physical_quantities}, with columns corresponding to the unigrid evolution, AMR with one refinement level, and AMR with two refinement levels.

During the interaction phase, the energy components exhibit large, correlated oscillations associated with wave interference and the redistribution of kinetic and gravitational energy. Peaks in $\rho_{\max}$ coincide with core overlap, while the virial parameter $\eta$ departs significantly from zero, reflecting the strongly dynamical regime. At later times, $\eta$ relaxes toward bounded oscillations, indicating the formation of a virialized remnant.

The total mass remains well conserved throughout the evolution, with a small and controlled decrease attributable to Kreiss--Oliger dissipation, satisfying $0.996 \leq M/M(0) \leq 1$. The diagnostic behavior is consistent across all configurations, demonstrating that adaptive refinement preserves global conservation properties even in strongly nonlinear regimes.

This test demonstrates that the AMR GPP solver robustly captures nonlinear multi-core interactions, interference dynamics, and relaxation processes, while preserving consistency and conservation across refinement levels.

\subsection{Merger of two line vortices}
\label{sec:vortex-collision}

{\it Purpose.}
This test extends the binary merger benchmark to configurations containing topological defects and assesses the ability of the AMR GPP solver to evolve strongly nonlinear vortex--vortex interactions. In contrast with the solitonic merger test, this problem probes the robustness of the numerical scheme in the presence of phase singularities, vorticity transport, and complex interference patterns. It provides a stringent test of refinement consistency and Poisson coupling in a regime where both density and phase dynamics are essential.

\medskip

{\it Initial conditions.}
The initial configuration consists of two identical stationary line vortices aligned with the $z$ axis, constructed following Appendix~\ref{app:line-vortices}. Each vortex has winding number $m=1$ and total mass $M \approx 25$ in dimensionless units. The vortices are placed symmetrically in the $(x,y)$ plane at
\[
\vec{x}_1 = (-8,-8,0), \qquad
\vec{x}_2 = (+8,+8,0),
\]
and assigned equal and opposite velocities along the $x$ direction,
\[
\vec{v}_1 = (+0.2,\,0,\,0), \qquad
\vec{v}_2 = (-0.2,\,0,\,0),
\]
resulting in a finite impact parameter and nonzero orbital angular momentum about the $z$ axis.

The total wavefunction is initialized as the coherent superposition
\begin{equation}
\Psi(0,\mathbf{x}) =
\Psi_{\mathrm{vortex}}(|\vec{x}-\vec{x}_1|)\,e^{i\vec{v}_1\cdot\vec{x}}
+
\Psi_{\mathrm{vortex}}(|\vec{x}-\vec{x}_2|)\,e^{i\vec{v}_2\cdot\vec{x}},
\end{equation}
with repulsive self--interaction $g=1$.

\medskip

\begin{figure}
    \centering
    \includegraphics[width=8cm]{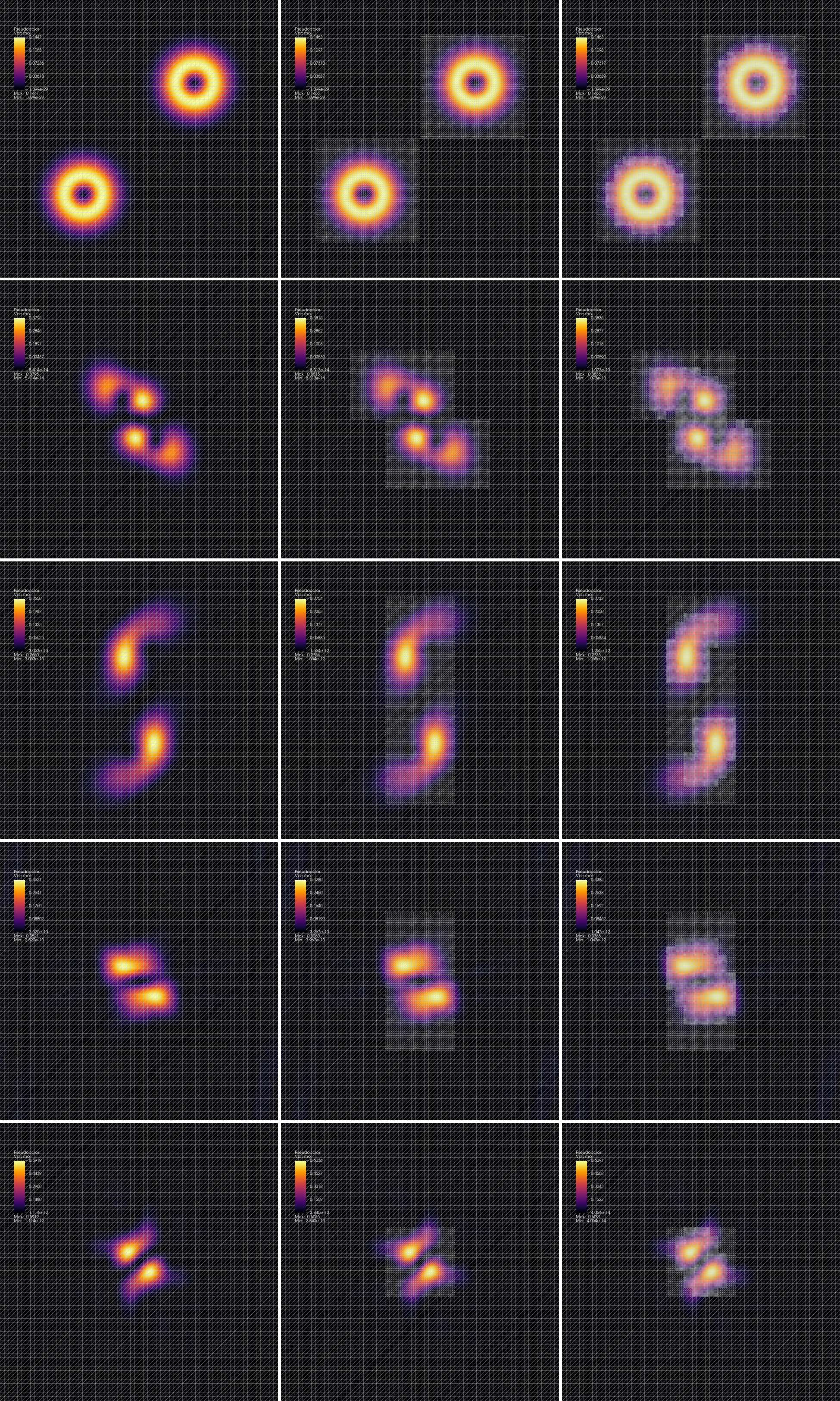}
    \caption{Merger of two line vortices under periodic boundary conditions for $g=1$. Density slices on the midplane $z=0$ are shown together with the computational mesh. Columns correspond to the unigrid evolution (left), AMR with one refinement level (center), and AMR with two refinement levels (right). Rows display snapshots at $t=0$, $40$, $80$, $120$, and $160$ (top to bottom), illustrating orbital motion, vortex--vortex interaction, and subsequent rearrangement of the density field.}
    \label{fig:vortex_binary_evolution}
\end{figure}

{\it Results.}
Figure~\ref{fig:vortex_binary_evolution} shows the evolution for three configurations: unigrid (left), AMR with one refinement level (center), and AMR with two refinement levels (right), with snapshots at $t=0$, $40$, $80$, $120$, and $160$.

Initially, the vortices orbit each other and generate complex interference patterns associated with the interaction of their phase singularities. During close encounters, the density field becomes highly distorted, producing filamentary and spiral structures. The dynamics are governed by the interplay between self--gravity and phase gradients, leading to strong redistribution of vorticity and angular momentum.

The AMR hierarchy dynamically tracks both the vortex cores and the surrounding interference structures, maintaining high resolution in regions where phase gradients are large. The morphology and evolution are consistent across all configurations, with no evidence of grid imprinting or numerical artifacts. This demonstrates that AMR accurately reproduces the unigrid solution even in the presence of topological defects and strongly nonlinear phase dynamics.

\medskip

\begin{figure}
    \centering
    \includegraphics[width=8cm]{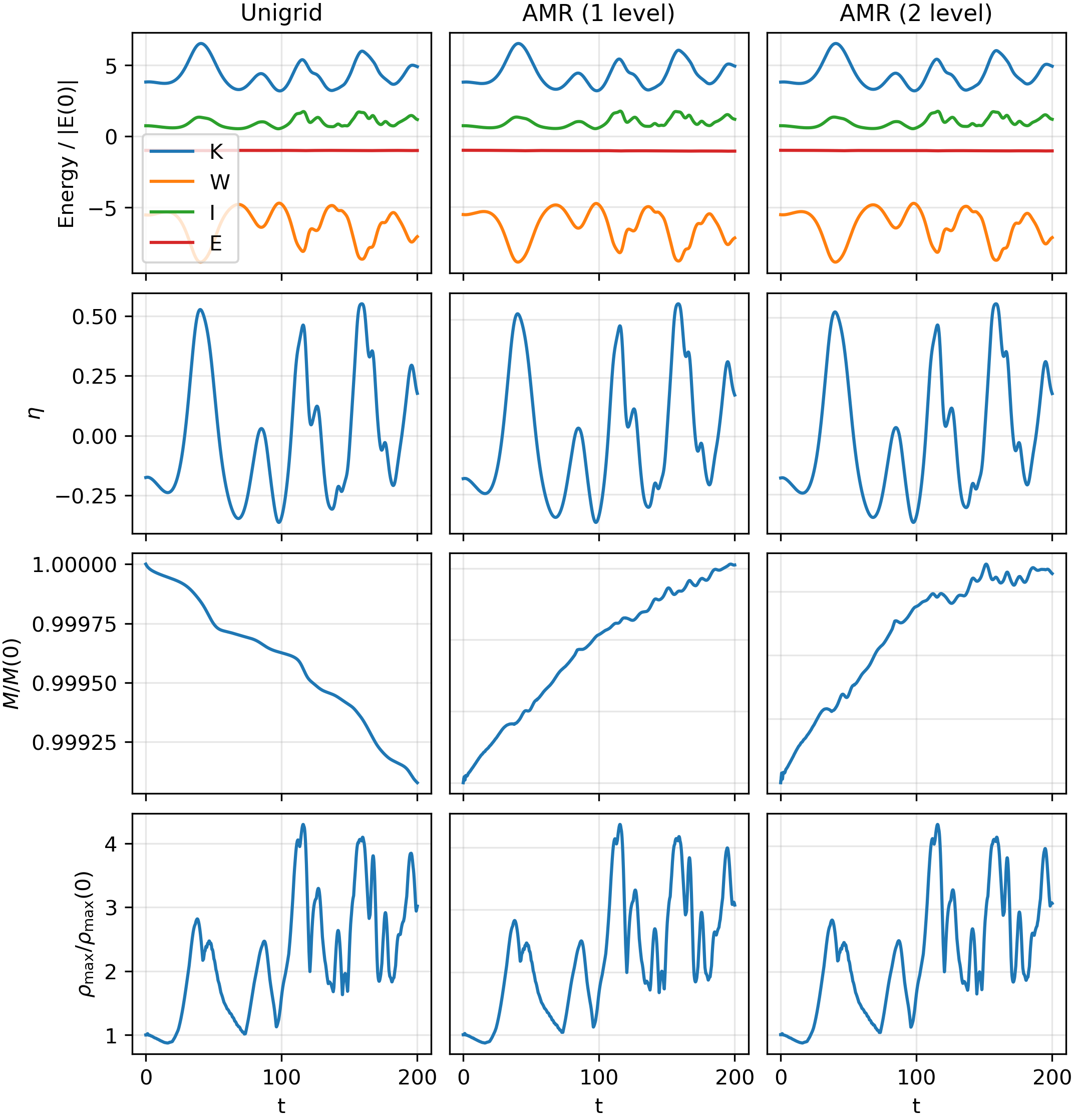}
    \caption{Global diagnostics for the merger of two line vortices with $g=1$. Columns correspond to the unigrid evolution (left), AMR with one refinement level (center), and AMR with two refinement levels (right). Rows show, from top to bottom: energy components normalized to $|E(0)|$, virial parameter $\eta$, total mass $M/M(0)$, and maximum density $\rho_{\max}/\rho_{\max}(0)$ as functions of time. The diagnostics capture the nonlinear interaction and subsequent relaxation, with consistent behavior across all configurations.}
    \label{fig:vortex_binary_diagnostics}
\end{figure}

{\it Diagnostics.}
Global diagnostics are shown in Fig.~\ref{fig:vortex_binary_diagnostics}, with columns corresponding to the unigrid evolution, AMR with one refinement level, and AMR with two refinement levels.

The energy components exhibit strong time--dependent oscillations reflecting wave interference, angular momentum redistribution, and energy exchange during close encounters. The virial parameter $\eta$ departs significantly from zero during the interaction phase and later settles into bounded oscillations, indicating relaxation toward a quasi--stationary state.

The maximum density $\rho_{\max}(t)$ displays pronounced peaks associated with vortex compression and core interaction, while the total mass remains well conserved, with a small and controlled decrease attributable to Kreiss--Oliger dissipation. The diagnostic behavior is consistent across all configurations, demonstrating that adaptive refinement preserves global conservation properties even in the presence of phase singularities.

This test demonstrates that the AMR GPP solver robustly captures vortex--dominated dynamics, accurately resolving phase singularities, interference patterns, and nonlinear interactions while maintaining consistency and conservation across refinement levels.

\subsection{Gravitational condensation of a bosonic cloud}
\label{sec:condensation}

{\it Purpose.}
This test verifies that the code reproduces the gravitational Bose--Einstein condensation process observed in self-gravitating FDM systems. Starting from a statistically homogeneous and isotropic bosonic cloud, the system is expected to undergo nonlinear relaxation and spontaneously form a compact solitonic core. The emergence of this attractor solution provides a stringent validation of the code in a fully nonlinear regime involving self-gravity, wave interference, and long-term relaxation dynamics.

{\it Initial conditions.} Following Ref.~\cite{Rusos2018}, the initial wavefunction is generated from a random field in momentum space. The configuration in physical space is obtained through the inverse Fourier transform

\begin{equation}
\Psi(0,\mathbf{x})
=
\mathcal{F}^{-1}
\!\left[
\hat{\Psi}(\mathbf{p})
\right],
\label{eq:initial_cloud}
\end{equation}

\noindent where the Fourier amplitudes are prescribed as

\begin{equation}
\hat\Psi(\mathbf p)
=
A
\exp\!\left(
-\frac{p^2}{2\sigma^2}
\right)
e^{i\Theta(\mathbf p)}.
\label{eq:momentum_distribution}
\end{equation}

The quantity $\Theta(\mathbf p)$ denotes a random phase uniformly distributed in the interval $[0,2\pi)$, $A$ is a normalization constant that fixes the total bosonic mass, and $\sigma=1$ specifies the momentum dispersion.

The simulations are performed in a periodic cubic domain of size $L=18$, with total mass $M=1005.3$, and vanishing self-interaction, $g=0$. Following Ref.~\cite{refId0}, all simulations are initialized on a uniform grid because the initial density field is statistically homogeneous and isotropic. We use $N=128$ grid points per spatial direction, corresponding to a spatial resolution $h=\frac{L}{N} = \frac{18}{128} = 0.140625$. The time step is chosen to satisfy the stability condition adopted in previous simulations of gravitational condensation~\cite{ChengNiemeyer2021},

\begin{equation}
\frac{\Delta t}{h^2}
<
\frac{1}{6\pi}.
\label{eq:stability_condensation}
\end{equation}

The relevance of this benchmark stems from the fact that gravitational condensation is now understood to be a generic property of self-gravitating bosonic systems.

The formation of compact solitonic cores from initially diffuse clouds was demonstrated in the kinetic regime by Levkov, Panin, and Tkachev~\cite{Rusos2018}. The resulting condensates are well described by the ground-state equilibrium solutions of the Schr\"odinger--Poisson system studied in Ref.~\cite{GuzmanUrena2004}.

Such solitonic configurations were first identified in cosmological FDM simulations by Schive et al.~\cite{Schive:2014dra}, who found that they naturally emerge at the centers of virialized halos. Subsequent studies showed that these cores continue to grow through ongoing mass accretion~\cite{ChengNiemeyer2021}.

More recently, black holes have been shown to accelerate the condensation process and act as condensation centers for FDM cores~\cite{palomareschavez2025}, while coupled boson--gas and fermion--gas systems have been found to exhibit the same attractor behavior~\cite{AlvarezGuzmanNiemeyer_2025}.

Therefore, reproducing gravitational condensation and the emergence of the corresponding solitonic attractor constitutes a demanding validation test for any numerical implementation of the Schr\"odinger--Poisson equations.

\medskip

\begin{figure}
    \centering
    \includegraphics[width=4cm]{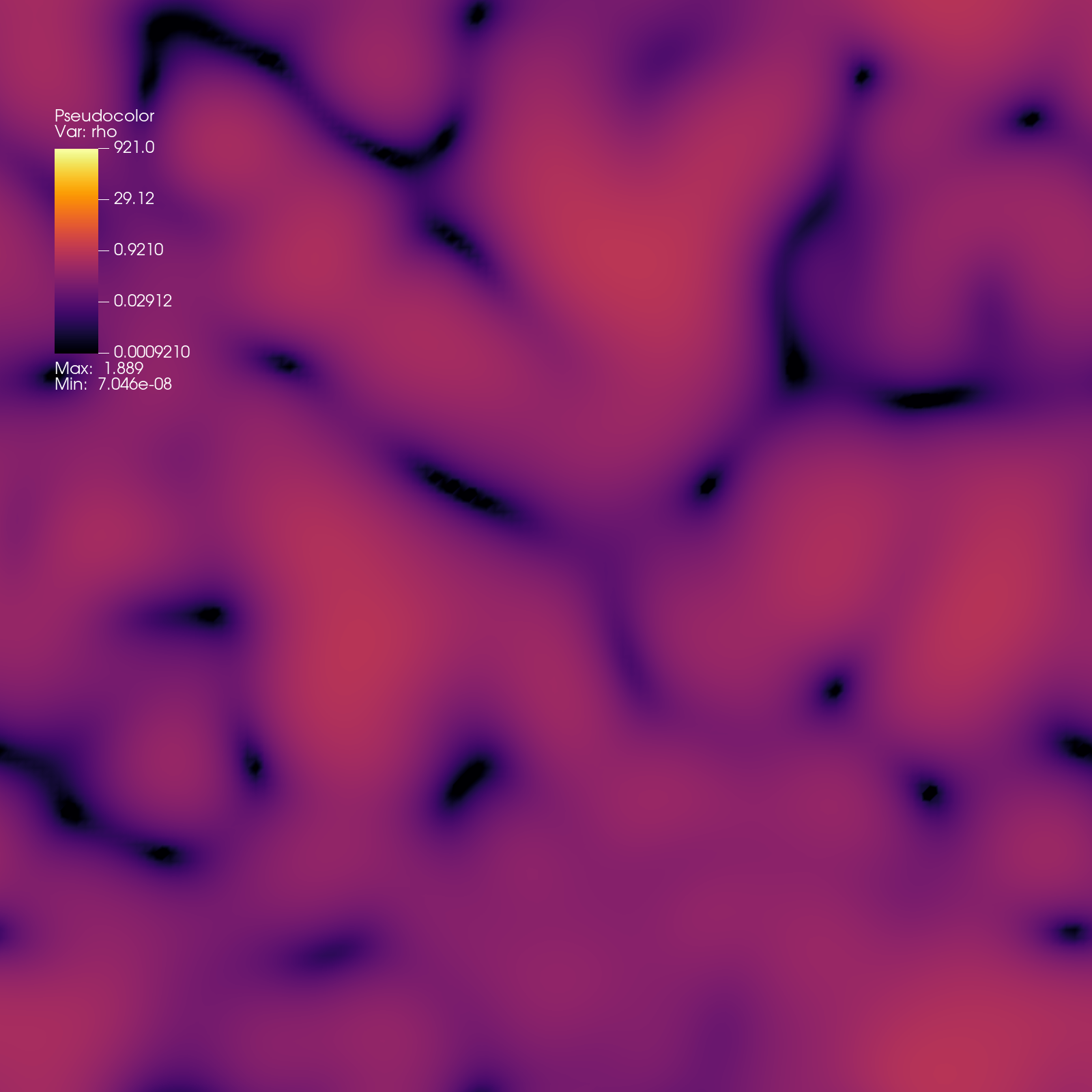}
    \includegraphics[width=4cm]{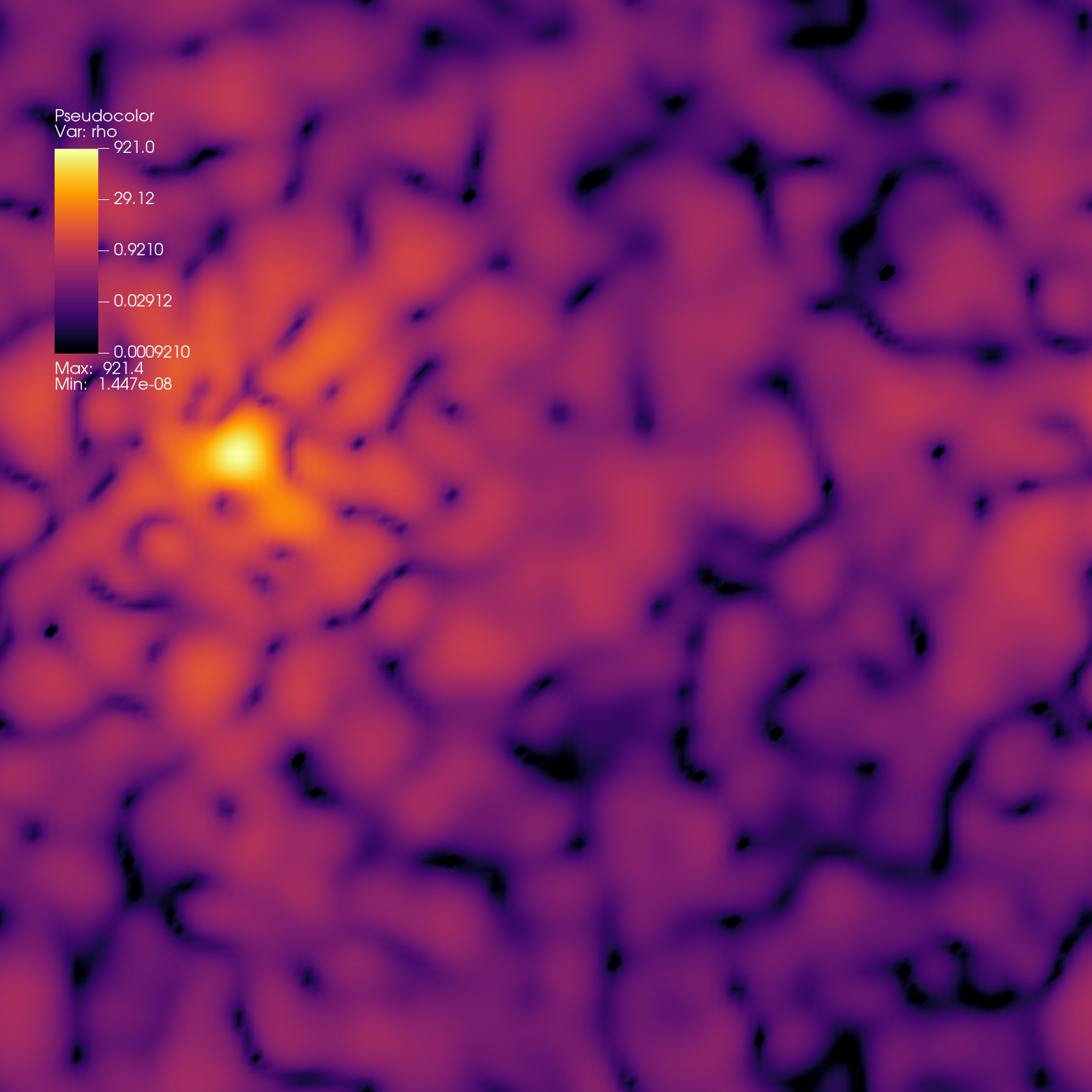}
    \caption{
    Gravitational condensation of a bosonic cloud in a periodic domain with $L=18$, $M=1005.3$, and $g=0$. Density slices on the midplane $z=7$ are shown at the initial time $t=0$ (left) and at the final time $t=100$ (right). The initially homogeneous and isotropic density field evolves through nonlinear gravitational relaxation and spontaneously develops a compact high-density condensate surrounded by a fluctuating halo, illustrating the onset of gravitational Bose--Einstein condensation.}
    \label{fig:condensation_evolution}
\end{figure}

{\it Results.}
Figure~\ref{fig:condensation_evolution} shows the evolution of the density field in the $(x,y)$ plane from the initial random configuration to the final state at $t=100$.

Initially, the density distribution is statistically homogeneous and isotropic, with fluctuations generated by the random phases in momentum space. As the system evolves, gravitational relaxation amplifies density perturbations and drives mass transport toward localized overdensities. By $t=100$, a compact high-density core has emerged from the initially diffuse cloud, while the surrounding medium develops a network of interference patterns and low-density regions characteristic of wave-supported self-gravitating dynamics.

The final configuration exhibits the qualitative features expected from gravitational Bose--Einstein condensation. A dominant central condensate forms spontaneously from the random initial state and coexists with an extended fluctuating halo. The peak density increases by nearly three orders of magnitude during the evolution, indicating the efficient transfer of mass into the condensed core.

\medskip

\medskip

\begin{figure}
    \centering
    \includegraphics[width=8cm]{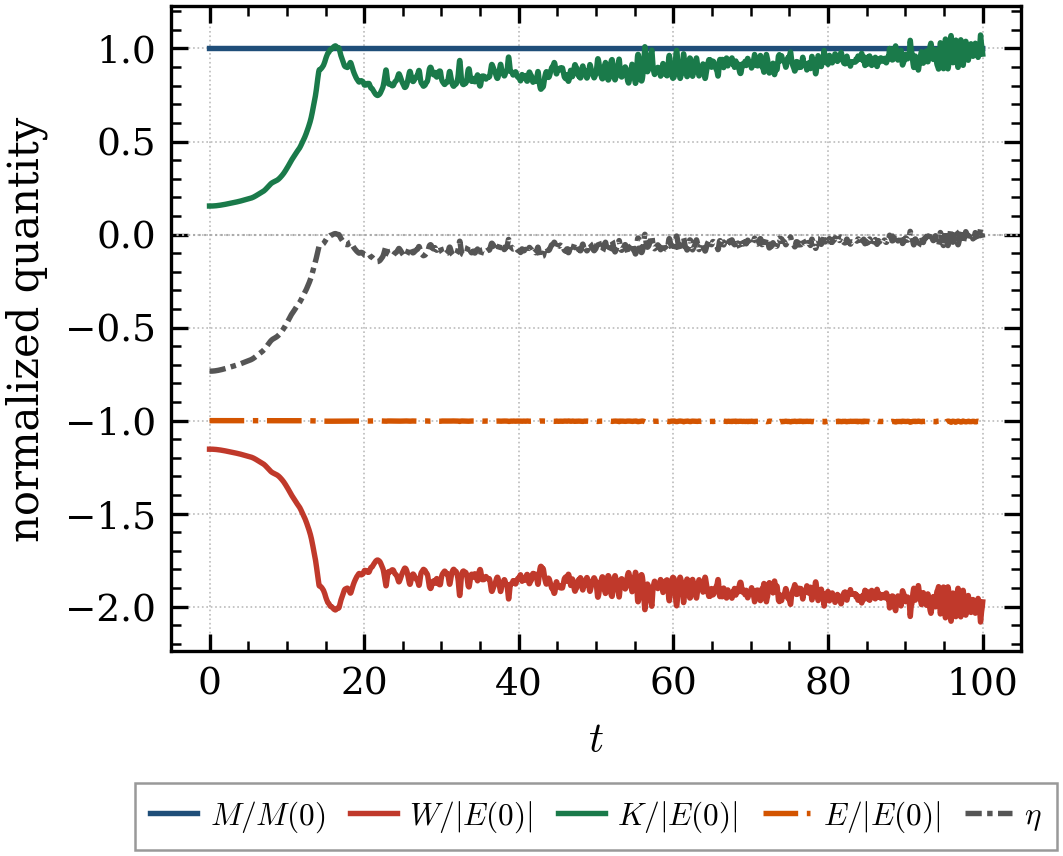}
    \caption{
    Global diagnostics for the gravitational condensation test. Shown are the total mass $M/M(0)$, total energy $E/|E(0)|$, kinetic energy $K/|E(0)|$, gravitational energy $W/|E(0)|$, and the virial parameter $\eta$ as functions of time. The diagnostics illustrate the relaxation of the initially random cloud toward a virialized condensed configuration while maintaining good conservation of mass and energy.
    }
    \label{fig:condensation_diagnostics}
\end{figure}

{\it Diagnostics.}
Global diagnostics are shown in Fig.~\ref{fig:condensation_diagnostics}. The displayed quantities include the total mass $M/M(0)$, the total energy $E/|E(0)|$, the kinetic energy $K/|E(0)|$, the gravitational energy $W/|E(0)|$, the virial parameter

\[
\eta=\frac{2K+W}{|W|},
\]

and the maximum density $\rho_{\max}(t)$.

The system begins from a non-equilibrium random configuration dominated by kinetic energy. During the early stages of the evolution, gravitational relaxation redistributes energy and drives mass toward localized overdensities. As a result, the kinetic energy increases while the gravitational energy becomes progressively more negative. The total energy remains nearly constant throughout the evolution, confirming that the dynamics are governed primarily by gravitational relaxation rather than numerical dissipation.

The virial parameter initially departs significantly from zero, reflecting the strongly non-equilibrium character of the initial cloud. Around $t\sim15$, the system undergoes a rapid relaxation episode associated with the onset of condensation. Subsequently, $\eta$ fluctuates around zero with small bounded oscillations, indicating that the configuration approaches a virialized quasi-stationary state.

\begin{figure}
    \centering
    \includegraphics[width=8cm]{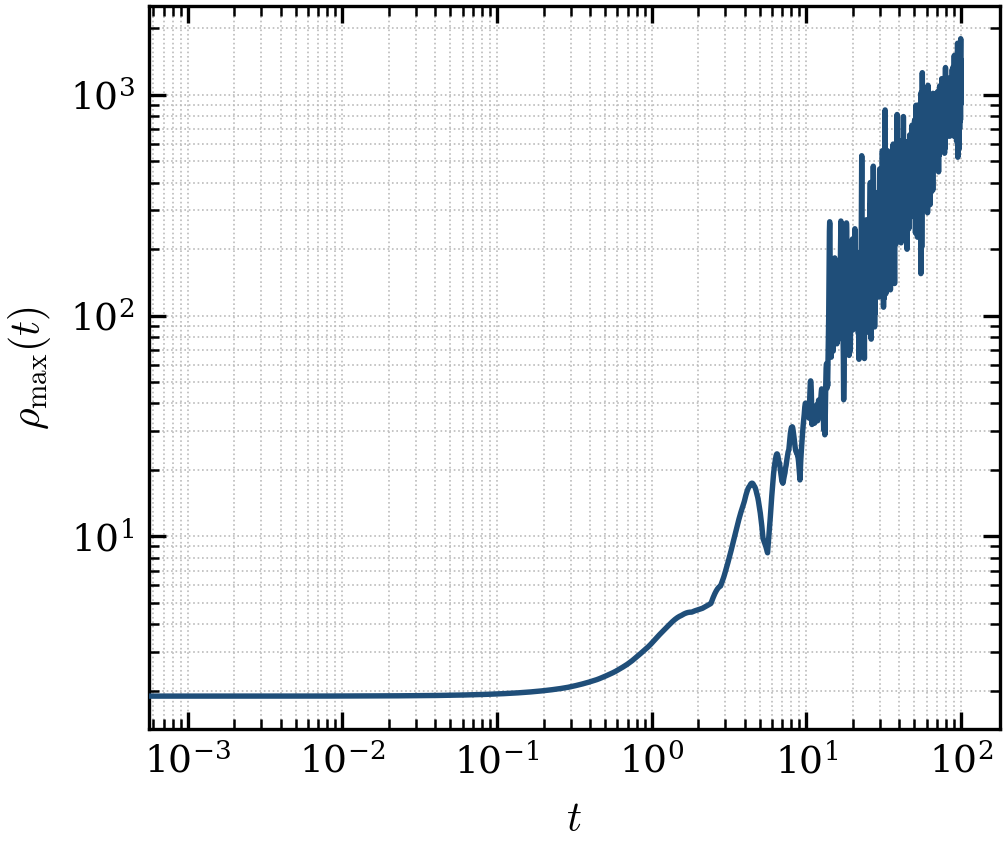}
    \caption{
    Evolution of the maximum density $\rho_{\max}(t)$ during gravitational condensation. After an initial phase of weak evolution, the maximum density increases by several orders of magnitude as a compact condensate forms. The continued growth at late times indicates ongoing accretion onto the central solitonic core.
    }
    \label{fig:condensation_rhomax}
\end{figure}

A particularly important diagnostic is the evolution of the maximum density shown in Fig.~\ref{fig:condensation_rhomax}. Starting from an approximately homogeneous distribution, $\rho_{\max}$ remains nearly constant during the initial phase and then grows rapidly by several orders of magnitude. This growth signals the emergence of a compact self-gravitating condensate and is consistent with the gravitational Bose--Einstein condensation mechanism reported in Ref.~\cite{Rusos2018}. Even after the initial collapse, the maximum density continues to increase, indicating ongoing accretion of material onto the central solitonic core, in agreement with the continuous core-growth scenario discussed in Refs.~\cite{ChengNiemeyer2021,palomareschavez2025}.

The emergence of a compact high-density core and its subsequent growth suggest that the condensate is approaching the well-known solitonic attractor of the Schr\"odinger--Poisson system. To verify this interpretation, we compute the spherically averaged density profile around the density maximum at several times during the evolution.

\begin{figure}
    \centering
    \includegraphics[width=8cm]{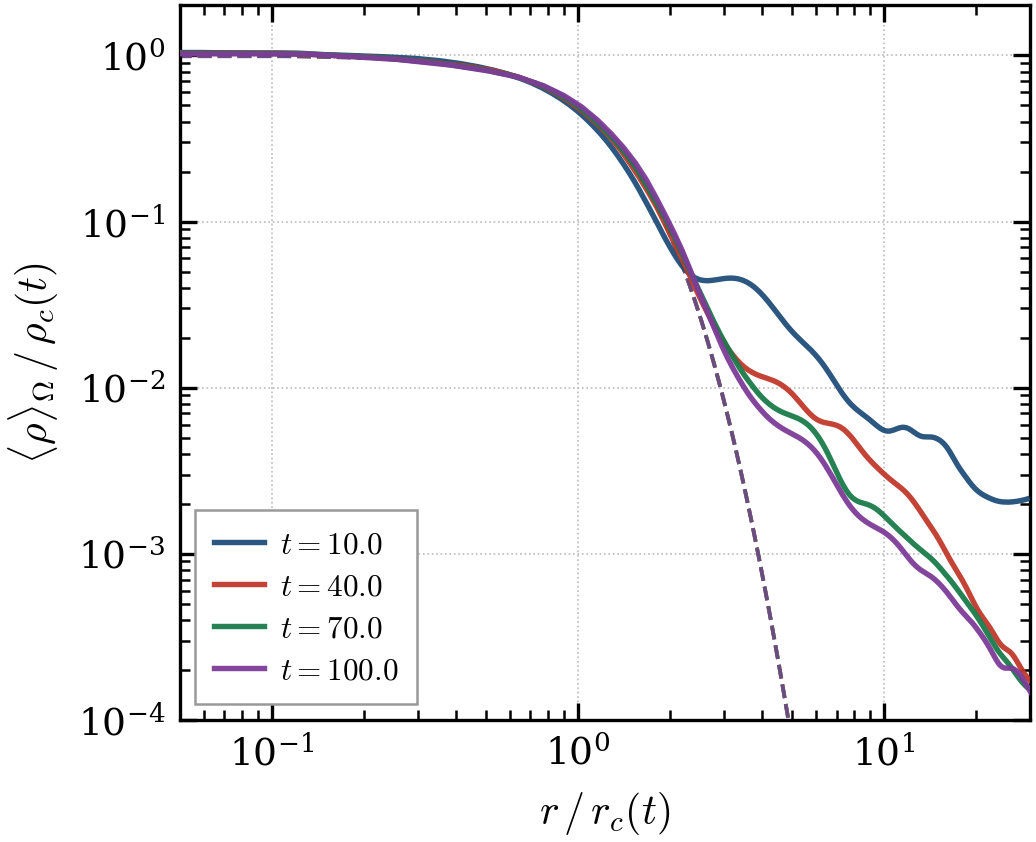}
    \caption{
    Spherically averaged density profiles centered on the condensate at different times during the evolution. Profiles are normalized by the instantaneous central density $\rho_c(t)$ and core radius $r_c(t)$. The dashed line corresponds to the empirical soliton profile (\ref{eq:core-g0}) of Schive et al.~\cite{Schive:2014dra}. As the system evolves, the inner density distribution converges toward the universal solitonic profile, confirming the formation of a self-gravitating condensate.
    }
    \label{fig:condensation_profile}
\end{figure}

Figure~\ref{fig:condensation_profile} shows the angularly averaged density profiles at $t=10$, $40$, $70$, and $100$, normalized by the central density $\rho_c(t)$ and core radius $r_c(t)$. As the condensation process proceeds, the inner structure converges toward a nearly universal profile. The late-time profiles become increasingly similar and are well described by the empirical soliton profile (\ref{eq:core-g0}) proposed by Schive et al.~\cite{Schive:2014dra}, shown by the dashed curve.

The agreement demonstrates that the condensed object formed through gravitational relaxation is not merely a transient overdensity, but corresponds to the same solitonic solution previously identified in cosmological FDM simulations and known to approximate the ground-state equilibrium configuration of the Schr\"odinger--Poisson system. This result confirms that the code successfully reproduces the gravitational Bose--Einstein condensation process and the emergence of the expected solitonic attractor.

\medskip

\section{Conclusions}
\label{sec:conclusions}

In this work, we have presented a numerical solver for the GPP system based on block--structured adaptive mesh refinement. The code is formulated in a fully time--dependent framework with periodic boundary conditions and fourth--order spatial discretization, and is designed to efficiently capture the multi--scale dynamics of self--gravitating bosonic matter.

The implementation has been validated through five increasingly demanding test problems in the nonlinear regime: the periodic advection of a boosted solitonic core, the transport of a boosted line vortex, the merger of two solitonic cores, and the merger of two line vortices, as well as the gravitational condensation of a random bosonic cloud into a solitonic core. These tests demonstrate that the solver accurately reproduces the unigrid solution across all configurations, while preserving global conservation properties, resolving interference patterns and phase singularities, and maintaining consistency across refinement levels, and reproducing the expected solitonic attractor of the Schr\"odinger--Poisson system.

Rather than aiming to outperform existing large--scale GPP solvers, the present implementation is intended as a complementary and independent framework. In this sense, it provides an additional tool for cross--validation of numerical results, helping to assess the robustness of physical conclusions against implementation details. At the same time, its relatively compact and modular design facilitates the rapid incorporation of new physical ingredients and numerical strategies.

The present framework therefore offers a flexible platform for future developments, including the exploration of alternative refinement criteria, coupling to additional matter components, and the study of more complex dynamical scenarios. The successful reproduction of gravitational Bose--Einstein condensation further demonstrates that the code accurately captures not only transient nonlinear dynamics but also long--term relaxation processes leading to self--gravitating equilibrium configurations. We expect that this code will serve both as a practical research tool and as a testbed for the development and validation of new models in simulations of self--gravitating quantum systems.

\begin{acknowledgments}
The author gratefully acknowledges Francisco S.~Guzm\'an and Flavio Rosales--Infante for valuable comments and insightful suggestions that improved the quality of this manuscript.
\end{acknowledgments}

\appendix
\section{Stationary solitonic solutions and empirical core profiles}
\label{app:stationary-cores}

The dimensionless GPP system,
Eqs.~\eqref{eq:gp-dimless}--\eqref{eq:poisson-dimless},
admits isolated, spherically symmetric stationary solutions of the form
\begin{equation}
    \Psi(t,\mathbf{x})=\psi(r)\,e^{-i\omega t},
\end{equation}
where the real radial profile $\psi(r)$ satisfies
\begin{align}
-\frac{1}{2r^2}\frac{d}{dr}\!\left(r^2\frac{d\psi}{dr}\right)
+V\,\psi+g\,\psi^3 &= \omega\,\psi,
\\
\frac{1}{r^2}\frac{d}{dr}\!\left(r^2\frac{dV}{dr}\right) &= \psi^2 .
\end{align}
Regular solutions obey $\psi'(0)=V'(0)=0$ and decay asymptotically as
$r\to\infty$.

\medskip

\textit{Empirical core profile.}
Stationary solutions are accurately described by the empirical density profile \cite{CarlosIvanFranciscoUniverse}

\begin{equation}
\rho_{\mathrm{core}}(r)
=
\rho_c
\left[
1+\left(2^{1/8}-1\right)
\left(\frac{r}{r_c}\right)^{2+\beta}
\right]^{-8},
\label{eq:core-general}
\end{equation}
where $\rho_c=\rho(0)$, and the parameters $r_c$ and $\beta$ encode the effect of
self--interaction. The non--interacting case ($g=0$) is recovered for $\beta=0$,
yielding the universal solitonic core profile
\begin{equation}
r_c \simeq 1.306\,\rho_c^{-1/4},
\qquad
\beta=0,
\label{eq:core-g0}
\end{equation}
which is known to be dynamically stable \citep{GuzmanUrena2004} and observed in
cosmological simulations \citep{Schive:2014dra}.

\medskip

\textit{Scaling relations.}
Exploiting the scaling symmetry of the GPP system, the core parameters depend on
the $\lambda$--invariant combination
\begin{equation}
    \alpha = g\,\rho_c^{1/2}.
\end{equation}
Fitting stationary numerical solutions yields the relations
\begin{align}
r_c
&=
1.306\,\rho_c^{-1/4}
\left(1+a_1\alpha+a_2\alpha^2\right),
\\
\beta
&=
b_1\alpha+b_2\alpha^2+b_3|\alpha|^{1/2},
\end{align}
with coefficients
$a_1=0.3681$, $a_2=0.0905$,
$b_1=0.2842$, $b_2=0.0845$, and $b_3=-0.0117$.
These expressions smoothly reduce to Eq.~\eqref{eq:core-g0} in the limit
$\alpha\to0$.

\medskip

\textit{Stability and critical coupling.}
The total core mass,
\begin{equation}
M_{\mathrm{core}}=\int_0^\infty \rho_{\mathrm{core}}(r)\,r^2\,dr,
\end{equation}
is a non--monotonic function of $\alpha$ and reaches a maximum at the critical
value
\begin{equation}
\alpha_c \simeq -0.7140.
\end{equation}
This value separates stable and unstable branches of stationary solutions:
configurations with $\alpha>\alpha_c$ are dynamically stable, while those with
$\alpha<\alpha_c$ are unstable. The critical value $\alpha_c$ coincides with the
maximum of $M_{\mathrm{core}}(\alpha)$ and agrees with the independent analysis \cite{Chen2021}.

In the present work, all equilibrium configurations used for validation tests
are chosen within the stable branch $\alpha>\alpha_c$, ensuring that the
observed dynamics originate from the numerical scheme and AMR treatment rather
than from intrinsic instabilities of the stationary solutions.

\section{Stationary line-vortex solutions}
\label{app:line-vortices}

The dimensionless GPP system,
Eqs.~\eqref{eq:gp-dimless}--\eqref{eq:poisson-dimless},
admits stationary solutions with nontrivial phase winding, corresponding to
quantized vortex lines. These configurations represent straight vortices aligned
with a fixed symmetry axis and constitute a natural extension of the
spherically symmetric solitonic solutions discussed in
Appendix~\ref{app:stationary-cores}.

\medskip

\textit{Axisymmetric vortex ansatz.}
We consider straight vortex lines oriented along the $z$ axis and adopt
cylindrical coordinates $(r_\perp,\varphi,z)$. Stationary vortex solutions are
sought in the form
\begin{equation}
\Psi(t,r_\perp,\varphi,z)
=
\psi(r_\perp,z)\,e^{i m \varphi}\,e^{-i\omega t},
\label{eq:vortex-ansatz}
\end{equation}
where $m\in\mathbb{Z}$ is the winding number and
$\psi(r_\perp,z)$ is a real amplitude function.
Substitution of Eq.~\eqref{eq:vortex-ansatz} into the GPP system yields the coupled
equations
\begin{align}
\left[-\frac{1}{2}
\left(
\frac{\partial^2}{\partial r_\perp^2}
+
\frac{1}{r_\perp}\frac{\partial}{\partial r_\perp}
-
\frac{m^2}{r_\perp^2}
+
\frac{\partial^2}{\partial z^2}
\right)
+
V
+
g\,\psi^2
\right]\psi&=
\omega\,\psi,
\label{eq:vortex-gpp}
\\
\frac{1}{r_\perp}\frac{\partial}{\partial r_\perp}
\left(
r_\perp\frac{\partial V}{\partial r_\perp}
\right)
+
\frac{\partial^2 V}{\partial z^2}
&=
\psi^2 ,
\label{eq:vortex-poisson}
\end{align}
where the centrifugal term proportional to $m^2/r_\perp^2$ enforces a density
depletion along the symmetry axis, producing a hollow vortex core and a phase
singularity.

\medskip

\textit{Boundary conditions and symmetries.}
Axial regularity requires the wavefunction to vanish at the symmetry axis,
\begin{equation}
\psi(0,z)=0,
\end{equation}
which guarantees a finite kinetic energy density for $m\neq0$.
Equatorial symmetry with respect to the plane $z=0$ is imposed by requiring
\begin{equation}
\frac{\partial\psi}{\partial z}(r_\perp,0)=0,
\end{equation}
so that the solution is invariant under $z\to -z$.

At large distances from the vortex core, the solution decays asymptotically,
\begin{equation}
\lim_{\sqrt{r_\perp^2+z^2}\to\infty}
\psi
=
\lim_{\sqrt{r_\perp^2+z^2}\to\infty}
\frac{\partial\psi}{\partial r_\perp}
=
\lim_{\sqrt{r_\perp^2+z^2}\to\infty}
\frac{\partial\psi}{\partial z}
=
0,
\end{equation}
ensuring normalizability of the stationary state.  
In the linear case ($g=0$), these conditions also guarantee uniqueness of the
solution.

For the gravitational potential, regularity and symmetry are imposed through
\begin{equation}
\frac{\partial V}{\partial r_\perp}(0,z)=0,
\qquad
\frac{\partial V}{\partial z}(r_\perp,0)=0,
\end{equation}
together with vanishing potential at large distances,
$V\to 0$ as $\sqrt{r_\perp^2+z^2}\to\infty$.

\medskip

\textit{Numerical construction.}
The stationary vortex solutions are obtained by solving
Eqs.~\eqref{eq:vortex-gpp}--\eqref{eq:vortex-poisson} as a nonlinear eigenvalue
problem for $(\psi,V,\omega)$ using an imaginary--time evolution method (ITEM).
At each iteration of the ITEM scheme, the Poisson equation is solved using a
successive over--relaxation (SOR) method subject to the boundary conditions
described above.

An initial guess consistent with the near--axis behavior
$\psi\propto r_\perp^{|m|}$ is evolved until convergence of the eigenfrequency
$\omega$ and the spatial profiles is achieved.

The resulting stationary vortex configurations are used as initial data for the
three--dimensional AMR simulations presented in the main text. The full complex
wavefunction is reconstructed according to
Eq.~\eqref{eq:vortex-ansatz} and embedded into the Cartesian grid without further
approximation.

\medskip

This appendix provides the minimal information required to reproduce the
stationary line--vortex initial data employed in the AMR validation tests, while
avoiding duplication of the detailed physical analysis presented in
Ref.~\citep{Alvarez_Rios_2025}.

\section{Long-term mass conservation of the boosted soliton test}
\label{app:longterm_mass}

To complement the validation presented in Sec.~\ref{sec:boosted-soliton}, we extended the boosted soliton benchmark to longer integration times and explored the dependence of mass conservation on the Kreiss--Oliger filtering strength.

The simulations were evolved up to
\[
t=1000
\]
code units and repeated for

\[
\epsilon_{\rm KO}
=
1.00,\qquad
1.25,\qquad
1.50,
\]

while keeping all remaining parameters fixed with respect to the original test.

Long-term mass conservation was quantified through

\[
\delta M(t)
=
\frac{M(t)-M(0)}{M(0)}.
\]

\begin{figure}
\centering
\includegraphics[width=8cm]{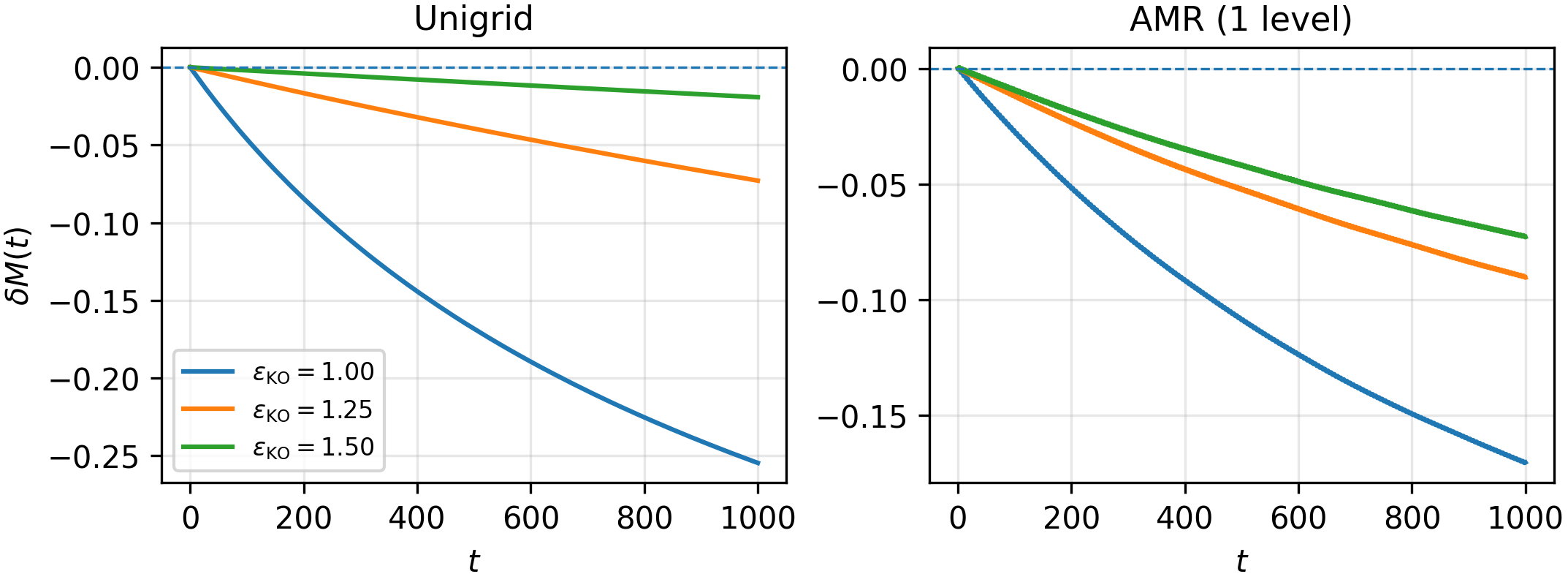}
\caption{
Long-term mass conservation for the boosted soliton test with $g=1$. Columns correspond to the unigrid evolution (left) and AMR with one refinement level (right). Curves show the relative mass variation $\delta M(t)$ for different values of the Kreiss--Oliger parameter $\epsilon_{\rm KO}$ as functions of time.
}
\label{fig:longterm_mass}
\end{figure}

Figure~\ref{fig:longterm_mass} shows that the accumulated mass drift decreases systematically as $\epsilon_{\rm KO}$ increases.

This behavior is consistent with the spectral interpretation discussed in Sec.~\ref{sec:filtering}. According to Eq.~(\ref{eq:filter}), increasing $\epsilon_{\rm KO}$ reduces the effective damping of modes near the grid cutoff and therefore weakens the action of the low-pass filter.

For both unigrid and AMR evolutions, weaker filtering improves long-term mass conservation while preserving stable evolution throughout the simulation.

An additional feature appears for
$\epsilon_{\rm KO}=1.00$,
where the AMR evolution exhibits smaller accumulated mass drift than the corresponding unigrid case.
This behavior is consistent with the larger local Nyquist frequency available in refined regions, which may reduce aliasing effects under strong filtering conditions.

For weaker filtering
($\epsilon_{\rm KO}=1.25$ and $1.50$),
the accumulated mass drift becomes larger in AMR than in the corresponding unigrid runs, suggesting that once aliasing is sufficiently controlled, inter-level synchronization effects become increasingly important contributors to long-term mass loss.

\medskip

\noindent
{\it Computational cost.} To quantify the computational overhead introduced by adaptive refinement, we additionally measured execution time and peak resident memory for the same boosted soliton benchmark.

Execution time is normalized to the corresponding unigrid run at the same base resolution to reduce dependence on hardware characteristics, while memory is reported in absolute units.

\begin{table}
\centering
\begin{tabular}{ccccc}
\hline
Base&
Configuration&
Effective Resolution&
Runtime&
Memory[MB]
\\
\hline

$64^3$
&
Unigrid
&
$64^3$
&
1.00
&
56.69
\\

$64^3$
&
AMR (1)
&
$128^3$
&
1.65
&
61.49
\\

$64^3$
&
AMR (2)
&
$256^3$
&
3.84
&
95.03
\\

\hline

$128^3$
&
Unigrid
&
$128^3$
&
1.00
&
227.30
\\

$128^3$
&
AMR (1)
&
$256^3$
&
1.42
&
246.47
\\

$128^3$
&
AMR (2)
&
$512^3$
&
1.98
&
293.80
\\

\hline
\end{tabular}
\caption{
Computational cost for the boosted soliton benchmark.
Runtime is normalized to the corresponding unigrid evolution at the same base resolution.
Memory corresponds to peak resident memory (RSS).
}
\label{tab:performance}
\end{table}

Table~\ref{tab:performance} shows that adaptive refinement increases computational cost sublinearly with respect to the increase in effective local resolution.

For the $64^3$ base grid, two refinement levels provide an effective local resolution equivalent to $256^3$ while increasing memory usage from $56.7$ MB to only $95.0$ MB.

Similarly, for the $128^3$ base grid, two refinement levels reach an effective local resolution of $512^3$ while increasing memory consumption from $227.3$ MB to $293.8$ MB and requiring less than twice the execution time of the corresponding unigrid evolution.

These results illustrate the ability of AMR to concentrate computational resources in dynamically relevant regions while avoiding the cost associated with globally refined meshes.

% -------------------------------------------------------
% -----     REFERENCES     ----------
% -------------------------------------------------------
\bibliography{BECDM}

\end{document}